\documentclass[a4paper,11pt]{article}
\usepackage{jheppubMod}
\usepackage{graphicx,epstopdf}
\usepackage{amsmath,amssymb}
\usepackage{tikz}
\usetikzlibrary{shapes,arrows}

\usepackage{contour}
\usepackage{tikz-cd}
\usepackage{array}
\usepackage{multirow}
\usepackage{mathtools}
\usepackage{subcaption}
\usepackage{accents}

\usepackage{lineno}

\usepackage{natbib,bm}
\usepackage{slashed}    
\usepackage{hyperref}
\usepackage{breakurl}
\usepackage{multirow}
\usepackage{bbold}
\usepackage{listings}
\usepackage{booktabs}
\usepackage{xcolor}

\usepackage{soul}
 
\def\be{\begin{equation}}
\def\ee{\end{equation}}
\def\bea{\begin{eqnarray}}
\def\eea{\end{eqnarray}}
\def\ge{\mathrel{\raise.3ex\hbox{$>$\kern-.75em\lower1ex\hbox{$\sim$}}}}
\def\la{\mathrel{\raise.3ex\hbox{$<$\kern-.75em\lower1ex\hbox{$\sim$}}}}
\def\simgt{\mathrel{\raise.3ex\hbox{$>$\kern-.75em\lower1ex\hbox{$\sim$}}}}
\def\simlt{\mathrel{\raise.3ex\hbox{$<$\kern-.75em\lower1ex\hbox{$\sim$}}}}
\newcommand{\mzh}{m_{V_h}}
\newcommand{\mzl}{m_{V_l}}
\newcommand{\gch}{g_{\chi h}}
\newcommand{\gcl}{g_{\chi l}}
\newcommand{\gql}{g_{Ql}}
\newcommand{\gqh}{g_{Qh}}

\newcommand{\gl}{g_l}
\newcommand{\gh}{g_h}

\newcommand{\fr}[2]{\frac{#1}{#2}}

\newcommand{\nc}{\newcommand}

\nc{\gone}{\bar g_{\pi NN}^{(1)}}
\nc{\gzero}{\bar g_{\pi NN}^{(0)}}
\nc{\al}{\alpha}
\nc{\ga}{\gamma}
\nc{\de}{\delta}
\nc{\ep}{\epsilon}
\nc{\ze}{\zeta}
\nc{\et}{\eta}
\nc{\ka}{\kappa}
\nc{\rh}{\rho}
\nc{\si}{\sigma}
\nc{\ta}{\tau}
\nc{\up}{\upsilon}
\nc{\ph}{\phi}
\nc{\ch}{\chi}
\nc{\ps}{\psi}
\nc{\om}{\omega}
\nc{\Ga}{\Gamma}
\nc{\De}{\Delta}
\nc{\La}{\Lambda}
\nc{\Si}{\Sigma}
\nc{\Up}{\Upsilon}
\nc{\Ph}{\Phi}
\nc{\Ps}{\Psi}
\nc{\Om}{\Omega}
\nc{\ptl}{\partial}
\nc{\del}{\nabla}
\nc{\ov}{\overline}
\nc{\us}{U(1)$_S$}

\mathchardef\mhyphen="2D

\def\beq{\begin{equation}}
\def\eeq{\end{equation}}
\def\bmat{\begin{displaymath}}
\def\emat{\end{displaymath}}
\def\bear{\begin{eqnarray}}
\def\eear{\end{eqnarray}}
\def\ba{\begin{eqnarray}}
\def\ea{\end{eqnarray}}
\def\bery{\begin{array}}
\def\ery{\end{array}}
\def\bit{\begin{itemize}}
\def\eit{\end{itemize}}
\def\ben{\begin{enumerate}}
\def\een{\end{enumerate}}
\def\btab{\begin{tabular}}
\def\etab{\end{tabular}}
\def\btbl{\begin{table}}
\def\etbl{\end{table}}
\def\bfig{\begin{figure}[htb]}
\def\efig{\end{figure}}
\def\bpic{\begin{picture}}
\def\epic{\end{picture}}

\DeclareMathOperator\erf{erf}

\def\ga{\mathrel{\raise.3ex\hbox{$>$\kern-.75em\lower1ex\hbox{$\sim$}}}}
\def\la{\mathrel{\raise.3ex\hbox{$<$\kern-.75em\lower1ex\hbox{$\sim$}}}}
\def\gappeq{\mathrel{\rlap {\raise.5ex\hbox{$>$}}
{\lower.5ex\hbox{$\sim$}}}}
\def\lappeq{\mathrel{\rlap{\raise.5ex\hbox{$<$}}
{\lower.5ex\hbox{$\sim$}}}}

\def\gyr{{\rm \, G\kern-0.125em yr}}
\def\mev{{\rm \, Me\kern-0.125em V}}
\def\gev{{\rm \, Ge\kern-0.125em V}}
\def\tev{{\rm \, Te\kern-0.125em V}}

\newlength{\dhatheight}
\newcommand{\doublehat}[1]{%
    \settoheight{\dhatheight}{\ensuremath{\hat{#1}}}%
    \addtolength{\dhatheight}{-0.35ex}%
    \hat{\vphantom{\rule{1pt}{\dhatheight}}%
    \smash{\hat{#1}}}}
\renewcommand{\d}[1]{\ensuremath{\operatorname{d}\!{#1}}}
\newcommand\ddfrac[2]{\frac{\displaystyle #1}{\displaystyle #2}}

\definecolor{TODO}{HTML}{A6262C}

\AtBeginDocument{
\heavyrulewidth=.08em
\lightrulewidth=.05em
\cmidrulewidth=.03em
\belowrulesep=.65ex
\belowbottomsep=0pt
\aboverulesep=.4ex
\abovetopsep=0pt
\cmidrulesep=\doublerulesep
\cmidrulekern=.5em
\defaultaddspace=.5em
}

\begin{document}
\title{Dark biportals at direct detection}
\author[a,d]{Leonie Einfalt,}
\author[b]{Suchita Kulkarni,}
\author[c]{Massimiliano Procura,}
\author[a,d]{Florian Reindl}
\affiliation[a]{Institute of High Energy Physics, Austrian
     Academy of Sciences, Nikolsdorfergasse 18, 1050 Vienna,
     Austria}
\affiliation[b]{Institute of Physics, NAWI Graz, University of Graz, Universit\"atsplatz 5, A-8010 Graz, Austria}

\affiliation[c]{University of Vienna, Faculty of Physics, Boltzmanngasse 5, 1090 Vienna,
     Austria}

\affiliation[d]{Technische Universit\"at Wien, Atominstitut, Stadionallee 2, 1020 Vienna, Austria} 
\affiliation{Institute of High Energy Physics, Austrian
     Academy of Sciences, Nikolsdorfergasse 18, 1050 Vienna,
     Austria}
     
\emailAdd{leonie.einfalt@oeaw.ac.at}
\emailAdd{suchita.kulkarni@uni-graz.at}
\emailAdd{massimiliano.procura@univie.ac.at}
\emailAdd{florian.reindl@oeaw.ac.at}

\preprint{UWThPh 2021-15}

\abstract{
We present a study of the constraining power of low-threshold, high-resolution direct-detection experiments for a biportal simplified dark matter model. In the scenario we consider here, dark matter and Standard Model particles interact through two dark vector mediators --- one light and one heavy with respect to the momentum transfer in the experiment. Interference effects at the level of scattering amplitudes can lead to novel marked shape features in the differential recoil spectra, which are best exploited by high-resolution, low-threshold experiments. We identify the region in parameters space for our model where such effects are dominant and show that composite-target experiments with large atomic mass differences are ideal to explore these scenarios. We develop a profile likelihood approach to analyze presently available and future data. Using published results by the CRESST-III experiment and projections of future sensitivities for the COSINUS experiment, we constrain the parameter space in our model, thereby showing the potential of such an analysis on a class of dark matter models which exhibit non-standard features in the recoil spectra.}
\maketitle
\section{Introduction}
\label{sec:intro}

An important avenue in the search for particle dark matter (DM) is direct detection, which aims to reveal interactions between DM and Standard Model (SM) fields in ultra-sensitive, Earth-bound detectors. Traditionally, these experiments have focused on measuring elastic scattering of nonrelativistic DM particles off atomic nuclei, with the dominant background typically arising from electron recoils. The key observable to identify a possible DM signal in these analyses is the differential nuclear recoil spectrum, whose shape is dictated by underlying particle physics models.

The interactions underlying scattering processes involved in these direct DM searches can be well approximated by effective contact operators if the mediator fields are sufficiently heavy compared to the typical momentum transfer~\cite{Fan:2010gt,Dent:2015zpa,Anand:2013yka,Fitzpatrick:2012ix}, which is of the order of 1--100 keV. If this is not the case, the expected changes in the shape of the recoil spectrum make low-threshold experiments respond better to the interaction mechanism~\cite{Fornengo:2011sz} (see also Refs. ~\cite{Li:2014vza,DelNobile:2015uua}).

In the last years, enormous technological progress has been made in lowering the nuclear recoil threshold of direct DM detection. A recent compilation of current results may be found, for instance, in Ref. \cite{billard_direct_2021}. In order to better understand the reach of these experiments, it is thus interesting and timely to analyze DM models that predict nonstandard features of the shape of the recoil spectrum which these low-threshold experiments might be sensitive to. Such analyses not only help demonstrate the importance of these experiments but may also point to previously unexplored signatures, which could be investigated with improved detector design. Motivated by these considerations, in this paper we consider the case study of a biportal simplified model of DM-quark interactions and focus on two cryogenic experiments with composite targets, namely, CRESST~\cite{CRESST2019_firstresults,CRESST:2017fmc,CRESST:2015djg,CRESST:2015txj}, whose detector material is calcium tungstate (CaWO\textsubscript{4}) and which features a low threshold, and the upcoming COSINUS experiment~\cite{Angloher:2016ooq}, which uses sodium iodide (NaI). The choice of COSINUS is motivated by the large mass difference between the target elements, which (as we demonstrate later) is important for probing our scenarios.

The very concept of DM-quark interaction via multiple mediators is not new and has been previously studied in the literature; see, e.g., Ref.~\cite{Arcadi:2017kky} and references therein. For example, in supersymmetric gauge theories, DM-nucleon scattering can receive contributions from both Higgs and squark exchanges. Models with both a scalar and a $Z'$ portal to the dark sector naturally arise in extensions of the SM gauge group involving a new $U(1)'$~\cite{Duerr:2016tmh,Ghorbani:2015baa}. Models with two vector mediators were studied in Ref.~\cite{Liu:2019iik} in the context of DM indirect detection and in Ref.~\cite{Cai:2021evx}. For the case of two scalar mediators, interference effects on LHC searches were studied in Ref.~\cite{Ko:2016ybp}, while direct-detection and associated flavor-physics constraints were discussed in  Refs.~\cite{Bell:2017rgi,Bell:2018zra}. A Higgs-leptoquark portal to dark matter was the subject of Ref.~\cite{Mohamadnejad:2019wqb}. Two-scalar-mediator models were also explored in experimental searches at the LHC, as in Ref.~\cite{ATLAS:2019wdu}. The phenomenology of the two-Higgs-doublet plus pseudoscalar model was studied in Refs.~\cite{Bauer:2017ota,LHCDarkMatterWorkingGroup:2018ufk}, and more general scenarios of multi-pseudoscalar-mediated DM were recently analyzed in Ref.~\cite{Banerjee:2021hfo}, with the goal of overcoming inherent limitations of the approach based on simplified models with a single mediator.
Multiple DM-quark interactions have also been studied in the framework of an effective field theory (EFT) taking into account interference effects~\cite{Catena:2015uua}.

Our focus here is on the direct-detection phenomenology of a simplified model with two dark vector mediators, $V_l$ and $V_h$, coupling to both DM and SM quarks. Crucially, a large hierarchy between the masses of these mediators is assumed, namely, $\mzl^2\leq q^2$ and $\mzh^2\gg q^2$, where $q^2$ denotes the squared momentum transfer. The resulting interference effects at the level of scattering amplitudes lead to interesting novel features in this model like sharp dips in the recoil spectra.
Our analysis demonstrates the importance of low-threshold, high-resolution experiments in possibly detecting these features. Novel aspects of our study include the application of a profile likelihood approach to determine exclusion limits on the model parameters. For this purpose, we make use of the public CRESST-III data release, as well as the future sensitivity of the COSINUS experiment and study the target dependence of the recoil spectra in our model setup.

The paper is organized as follows. In Sec.~\ref{sec:multi_port} we introduce our vector biportal model, derive the expression for the differential recoil spectrum from the tree-level DM-nucleus cross section, discuss its characteristic features due to constructive and destructive interference and assess the effects of detector-specific quantities such as threshold and resolution for benchmark values of the model parameters. In Sec.~\ref{sec:Analysissetup} we discuss the likelihood formalism we applied to use published real data from the CRESST-III experiment and the projected sensitivity of the COSINUS experiment. Exclusions limits for the model parameters are presented and discussed in Sec.~\ref{sec:results} and Sec.~\ref{sec:conclusions} contains our conclusions.
\section{Vector biportal model at direct-detection experiments}
\label{sec:multi_port}
\begin{figure} [tb]
    \centering
    \begin{subfigure}[t]{0.8\textwidth}
        \begin{center}
      \includegraphics[width=\columnwidth]{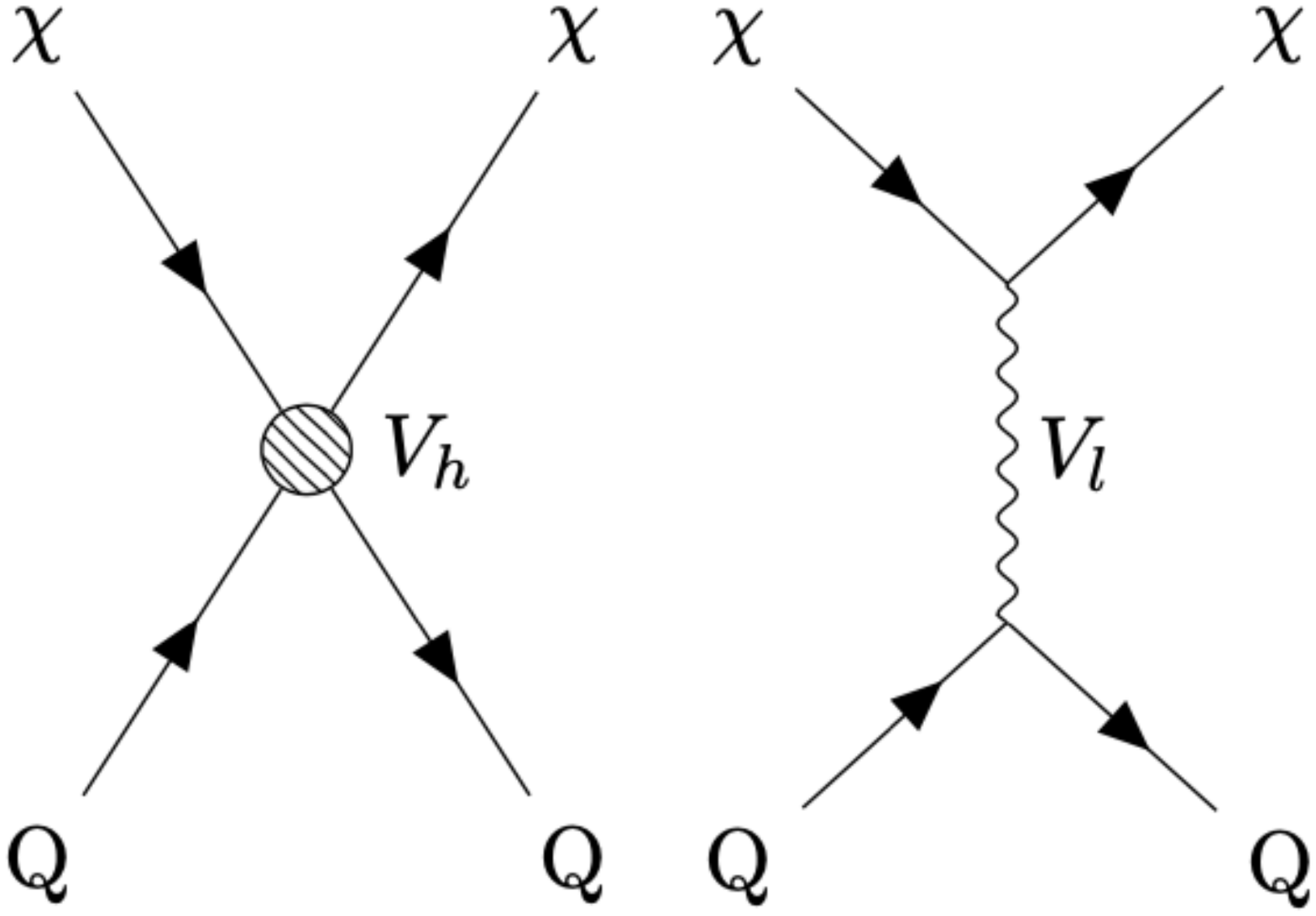}
              \end{center}
    \end{subfigure}
    \caption[Feynman diagram two vector mediators]{Feynman diagrams for a DM--quark scattering process at the lowest perturbative order within our model, with one heavy ($V_h$) and one light mediator ($V_l$).}
    \label{fig:feynman}
\end{figure}
The simplified model setup we consider consists of a Dirac DM particle $\chi$ interacting in a flavor-independent way with the SM quarks via a light and a heavy vector mediator ($V_l$ and $V_h$, respectively), with different couplings.  
The corresponding interaction terms in the relativistic Lagrangian amount to
\begin{equation}
    \label{eq:lagrangian}
    \mathcal{L}_{\rm int} \supset  \gql\, V_{l_{\mu}}\bar{Q}\gamma^{\mu} Q + \gcl\, V_{l_{\mu}}\bar{\chi}\gamma^{\mu} \chi +
    \gqh\, V_{h_{\mu}}\bar{Q}\gamma^{\mu} Q + \gch\, V_{h_{\mu}}\bar{\chi}\gamma^{\mu} \chi,
\end{equation}
where the SM quarks are denoted by $Q$ in order to avoid confusion with the momentum transfer $q$.
The dimensionless constants $\gql,\,\gcl$ denote the couplings between $V_l$ and the quarks, and $V_l$ and the DM particle, respectively. Analogous definitions hold for $\gqh, \gch$ in the case of $V_h$.

The Lagrangian above gives a general, effective, low-energy description of the interactions relevant for our analysis and is representative of a whole class of UV-complete theories featuring two vector mediators with a large mass hierarchy and flavor-independent vector couplings. This theoretical scenario, as we are going to show, is characterized by the fact that the resulting nuclear recoil spectra show nonstandard features like marked dips, due to interference effects. We point out that in the future it will also be interesting to investigate the modifications to these features of the recoil spectra corresponding to a generalization of our biportal simplified model to include further coupling structures as well as flavor-dependent mediator couplings.

We first derive the differential DM-nucleus cross section in terms of  couplings and masses in our model. This leads to the expression for the recoil rate, which our numerical analysis is based on. The total unpolarized squared scattering amplitude from the diagrams shown in Fig.~\ref{fig:feynman} is given in the nonrelativistic limit by 

\begin{align}
   \langle |M|^2\rangle &= 36 m_{\chi}^2\,m_N^2 A^2  F^2(E_R)  \Big[\frac{\gch^2\gqh ^2}{\mzh^4}+\frac{\gcl^2\gql^2}{(2m_N\,E_R + \mzl^2)^2} + 
    2\frac{\gch \gqh}{\mzh^2} \frac{\gcl\gql}{(2m_N\,E_R+\mzl^2)} \Big].
    \label{eq:scatamp}
\end{align}

In Fig.~\ref{fig:feynman}, the contact interaction is generated by integrating out the heavy mediator field $V_h$.
The magnitude of the momentum transfer  amounts to $q \equiv \sqrt{\vec{q}^{\;2}} = \sqrt{2 m_N E_R}$. Here $A$ denotes the mass number of the target nucleus and $m_N$ is its mass. The structure of the nucleus at nonzero momentum transfer is approximated by the standard form factor $F^2(q)$.\footnote{A systematic formalism to go beyond this approximation, including all one- and two-body spin-independent WIMP-nucleon interactions up to third order in chiral EFT was proposed in Refs.~\cite{Hoferichter:2016nvd,Hoferichter:2018acd}.} In our numerical analysis, we use the functional form for $F(q)$ first proposed by Helm~\cite{FF_Helm}, with the nuclear radius parametrization by Lewin and Smith \cite{FF_Lewin}.

An essential feature of the model is the interference term arising in the squared scattering amplitude. This plays an important role in the resulting phenomenology, as we are going to discuss in detail. It is also noteworthy that such an interference term would also be present in the case of the analogous exchange of two scalar mediators \cite{Ko:2016ybp, Morgante:2018tiq, Bell:2017rgi,Bell:2018zra}.

For the light mediator we define an overall dimensionless coupling $g_l \equiv \gcl\gql$ since the tree-level squared amplitude depends on this product and not on the separate values of $\gcl$ and $\gql$. Also for the heavy mediator, since we are only sensitive to the ratio between the product of the couplings and the mediator mass squared, we introduce an effective coupling $g_h \equiv (\gch \gqh)/\mzh^2$ of mass dimension $-2$. The dependencies on the couplings are relevant when thinking about the complementarity of the constraints on the biportal models from collider experiments and relic density mechanisms.

From Eq.~(\ref{eq:scatamp}) the tree-level differential cross section for nonrelativistic kinematics is given by
\begin{equation}
\label{eq:diffcrosssection}
    \frac{\d \sigma}{\d E_R}=\frac{m_N}{2\pi v^2} \,F^2(E_R) \,9 A^2 \left[\gh^2+\frac{\gl^2}{(2m_N E_R+\mzl^2)^2}+ 2\frac{\gh\gl}{2m_NE_R+\mzl^2}\right]~,
\end{equation}
where the third term in brackets is the contribution from interference. For $g_l\,g_h > 0$ we have constructive interference, while for $g_l\,g_h < 0$ there is destructive interference. For a given target, the free parameters in this equation are $g_l$, $g_h$ and $\mzl$.

The observable of interest in direct-detection experiments is the differential nuclear recoil rate. As we will see later on, experiments with high resolution and corresponding low threshold are needed in order to investigate the full phenomenology of the differential recoil spectrum in our model. This is given as a function of the differential cross section above by
\begin{align}
\label{eq:rate_DM}
\left.\frac{\d R}{\d E_R}\right|_{\chi} &= \mathcal{N} \,{\frac{\rho_0}{m_N\,m_{\chi}}} \int_{v_{\rm min}}^{v_{\rm esc}}d^3v\,|\vec{v}| \,f(\vec{v})\,\frac{\d \sigma}{\d E_R}
\end{align}
where $\rho_0$ is the local DM density, $\mathcal{N}$ is the (fiducial) mass of the detector, $|\vec{v}|$ is the DM speed in the lab frame, $v_{\rm min}(E_R)$ is the minimum DM speed required to generate a nuclear recoil with energy $E_R$ for an elastic collision and $v_{\rm esc}$ is the Galactic escape speed. The DM velocity distribution, denoted by $f(\vec{v})$, is in principle modulated in time due to the Earth's motion around the Sun, but we average out this effect here as we consider measurement times of 1 year or longer. We assume a normalized isotropic Maxwell-Boltzmann distribution in the Galactic rest frame 

\begin{align}
f_{\rm gal}(\vec{v}_{\rm gal}) &= N_{\rm esc}
\left( \ddfrac{3}{\,2\pi\,\sigma_v^2}\right)^{3/2}\,\exp\left[\ \ddfrac{- 3 \vec{v}_{\rm gal}^{\,2}}{2\sigma_v^2}\right] \\
\intertext{with}
N_{\rm esc} &=  \left[ \erf\left(z\right) - \frac{2}{\sqrt{\pi}}z\exp(-z^2) \right]^{-1}
\end{align}
where $\sigma_v = \sqrt{\frac{3}{2}}v_\infty$ is the root-mean-square velocity, which is related to the asymptotic value of the rotational velocity $v_\infty$, and $z^2 = 3 v_{\rm esc}^2/(2\sigma_v^2)$ \cite{Donato_1998galacticrot}. The distribution $f(\vec{v})$ in the Earth's rest frame is obtained by applying a Galilean transformation. In our numerical analysis we take $\rho_0=0.3$~GeV $c^{-2}$ cm$^{-3}$ \cite{Holmbergdensity2_2000},  $v_\infty=220$ km s$^{-1}$, and set the Earth's average speed equal to $1.05\, v_\infty $ \cite{Donato_1998galacticrot}. We also take $v_{\rm esc} = 544$ km s$^{-1}$, which is in agreement with the bounds set in Refs.~\cite{1987escape,Deason_2019escape}.  In the following we will neglect uncertainties in the determination of the local DM density (see, e.g., Ref.~\cite{Read:2014qva} for a review) as well as in the DM velocity distribution since a treatment of these effects is outside the scope of this work. Finally, for detectors with composite targets, a weighted sum over all targets in Eq.~(\ref{eq:rate_DM}) needs to be performed.

\subsection{Features of the recoil spectra in the biportal model}
\label{subsec:Phenom. analysis}

We now illustrate and discuss the peculiar features of the differential recoil spectra in our biportal model as compared to the single-mediator case. For this purpose, in this section we fix the four parameters that are left free in Eq.~(\ref{eq:rate_DM} to the benchmark values listed in Table~\ref{tab:freeparameters} and explore the effects of the variations of the parameters around these values.

In the benchmark model the mass of the DM particle and the light mediator $V_l$ are set to 5 GeV and 10 MeV, respectively. The values of $g_l$ and the effective coupling $g_h$ are chosen such that differential recoil rates are compatible with current limits on the spin-independent DM-nucleon cross section~\cite{DarkSide_2018}. For a DM mass of $5\,\,\rm GeV$ with $\sigma_{SI}<10^{-42}\,\,\rm cm^2$ \cite{DarkSide_2018}, we estimate $g_l=\mathcal{O}(10^{-12})$ and $g_h=\mathcal{O}(10^{-14}\,\,\rm MeV^{-2})$ for an $\mzh$ of about $100 \,\,\rm{MeV}$. The benchmark values of the couplings for the light and heavy mediator are chosen to satisfy this constraint while aiming to maximize interference effects. 

\begin{table}[t]
    \small
    \centering
    \renewcommand{\arraystretch}{1.4}
    \begin{tabular}{|>{\centering\arraybackslash}m{2cm}|>{\centering\arraybackslash}m{6cm}|>{\centering\arraybackslash}m{3cm}|} 
     \hline
     Parameter & Description & Benchmark value  \\ 
     \hline\hline
        $m_\chi$ & DM particle mass &$ 5\,\,\rm GeV$ \\
        $\mzl$ & Light mediator mass &$ 10\,\,\rm MeV $ \\
        $g_l$ & Light mediator coupling &$ 2 \times 10^{-12} $ \\
        $g_h$ & Heavy mediator effective coupling & $10^{-14}\,\,\rm MeV^{-2} $ \\
     \hline
    \end{tabular}
    \caption{The four free parameters in the recoil spectrum for our biportal model and their benchmark values used in this section. The effects of varying $g_l$ and $g_h$ are discussed in the text and illustrated in the next figures.}
    \label{tab:freeparameters}
\end{table}  

We briefly comment on the relevance of collider constraints on our model parameters before proceeding further with our discussion.\footnote{In models with suitable additional DM-SM couplings, indirect-detection constraints can also play a role. Furthermore, anomaly cancellation in UV-complete theories with dark vector bosons associated with a new $U(1)'$ is known to require the existence of multiple dark fermions and, depending on the mass of these additional particles, one can expect new search avenues for such models~\cite{Cui:2017juz}.} Vector mediators coupling to SM fields are being extensively searched for, both in fixed-target and collider experiments. For very light mediators, constraints from astrophysical limits exist as well. However most studies based on laboratory experiments leave the mediator-DM coupling unrestricted, as they search for visible final states. Since what is relevant in our analysis is the product of DM-$V_l$ and SM-$V_l$ couplings, we can probe the parameter space where DM-$V_l$ is larger and thus can escape the collider constraints. Similar arguments apply to heavier mediators as well. The relevant searches for heavier mediators arise from fixed-target experiments as well as colliders. The relevant constraints depend on the mass of the mediator as well as the mediator branching ratio to specific final states. For example, if the heavy mediators have masses of the order of $100$ GeV, LHC constraints via resonance searches are applicable. In this case the mediator couplings to leptons are more strongly constrained than mediator couplings to quarks or DM. For mediators lighter than 100 GeV, other experimental constraints apply and depending on the mediator branching ratio, the exact limits differ. However, at direct detection experiments mediators with masses larger than about 10 MeV are considered heavy\footnote{The exact mediator mass at which the momentum transfer at direct-detection experiments becomes irrelevant is dependent on the target and DM mass. In our analysis we take $\mzl \sim 10\, \,\rm{MeV}$, which falls within the light-mediator regime for CRESST and COSINUS targets.} and hence the exact constraints become model dependent. 

\begin{figure}[tb]
    \centering
    \begin{subfigure}[t]{0.49\textwidth}
        \includegraphics[width=\columnwidth]{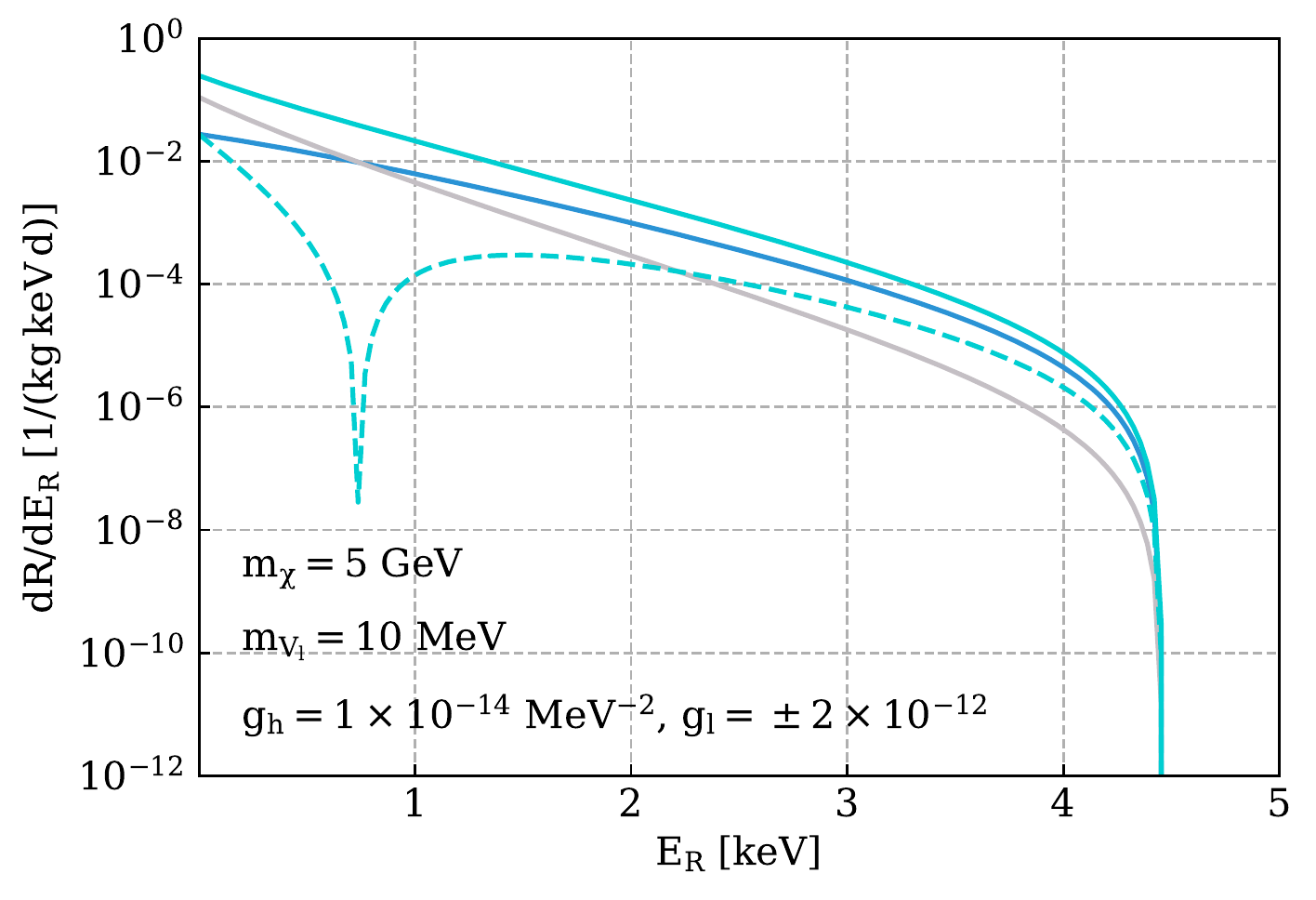}
        \subcaption{}
        \label{subfig:biportal1a}
    \end{subfigure}
    \begin{subfigure}[t]{0.49\textwidth}
        \includegraphics[width=\columnwidth]{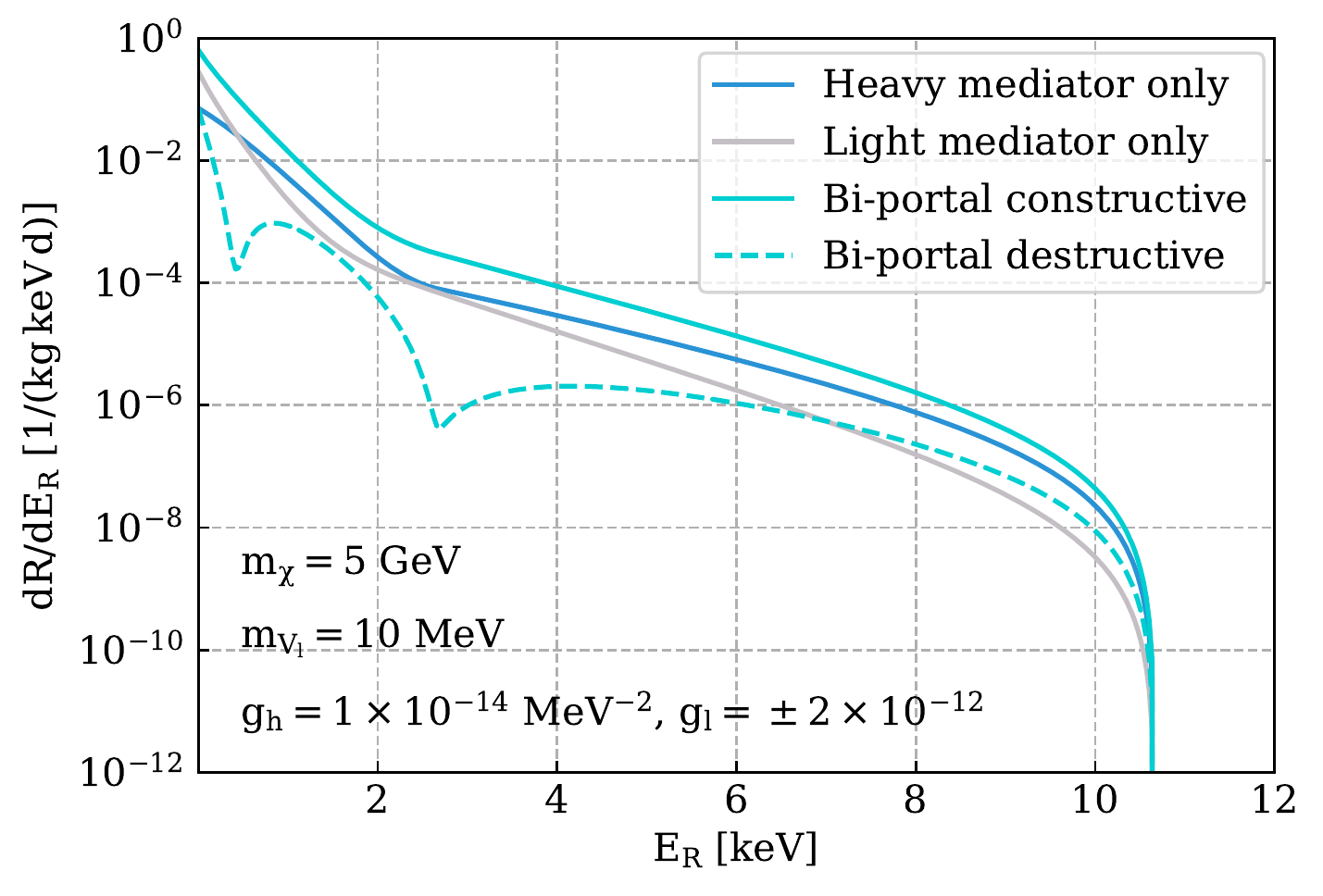}
        \subcaption{}
        \label{subfig:biportal1b}
    \end{subfigure}
 \caption{Differential event rate for the biportal mediator model on Ge (a) and NaI (b) targets. The light mediator mass is fixed at 10 MeV, the DM mass is fixed at 5 GeV, and the effective coupling of the heavy mediator at $1\times 10 ^{-14}\,\,\rm MeV^{-2}$. Effects of detector threshold and resolution are neglected here and will be considered in the next subsection. Differential event rates for a single light mediator model (grey lines), a single heavy mediator model (blue lines) model, and a biportal mediator model with constructive (turquoise solid lines) and destructive interference (turquoise dashed lines) are shown.}
\label{fig:biportal1}
\end{figure}

We now discuss the features of the differential nuclear recoil spectra in the benchmark vector biportal model.  The cases of a single-element target, namely germanium (Ge), and the composite target sodium iodide (NaI) are shown in Figs.~\ref{subfig:biportal1a} and \ref{subfig:biportal1b}, respectively, for both constructive (solid turquoise line) and destructive interference (dashed turquoise line). To compare the biportal model with single-mediator models, we also show the spectra for the heavy (blue line) and light mediator component (grey line). 

As expected, the constructive biportal model predicts an overall higher differential recoil rate, irrespective of the target. The shape is a nonlinear combination of the two individual single-mediator rates due to the presence of a non-negligible interference term. In the case of destructive interference, the situation becomes even more interesting. 

As shown in Fig.~\ref{subfig:biportal1a}, an overall lower rate is observed, but more interestingly a strong interference effect is present at around $E_R = 0.75\,\, \rm{keV}$. This dip appears precisely where the rates for the single heavy and single light mediators overlap and, due to the negative interference term, lead to a strong suppression. We stress that this dip is unique to the biportal model, as similar spectral features are not present in the case of interactions via one mediator. While the presence of this kink is a universal feature of biportal models, the exact location and depth depend on the target material as well as all of the free model parameters.

The impact of the interference term becomes even more apparent for composite targets. As shown in Fig.~\ref{subfig:biportal1b}, the biportal model with constructive interference (solid turquoise line) inherits a damped version of the kink present for the single heavy mediator case around a recoil energy of 2 keV. For NaI, this kink in the single heavy mediator spectrum marks the change of the dominant target which is iodine at low recoil energies and sodium at higher recoil energies. The recoil spectrum for the biportal model with destructive interference (dashed turquoise line) in Fig. \ref{subfig:biportal1b} now shows two interference effects. The first feature appears at rather low recoil energies, which is the iodine-dominated regime of the spectrum, and the other feature is present in the sodium-dominated regime. In the case of destructive interference, the biportal model can thus lead to multiple pronounced features in the expected recoil spectra of composite targets. However, these features are maximally enhanced when the various components of the target material vary strongly in their atomic mass/weight (e.g.,~Al$_2$O$_3$ only exhibits one spectral feature despite being a multielement target\footnote{The atomic weight of aluminium is about 27, while that of oxygen is 16.}). In order to probe the biportal model, especially the destructive case, composite targets with a pronounced mass difference between their elements are thus advantageous. 

\begin{figure} [tb]
    \centering
    \begin{subfigure}[t]{0.49\textwidth}
        \includegraphics[width=\columnwidth]{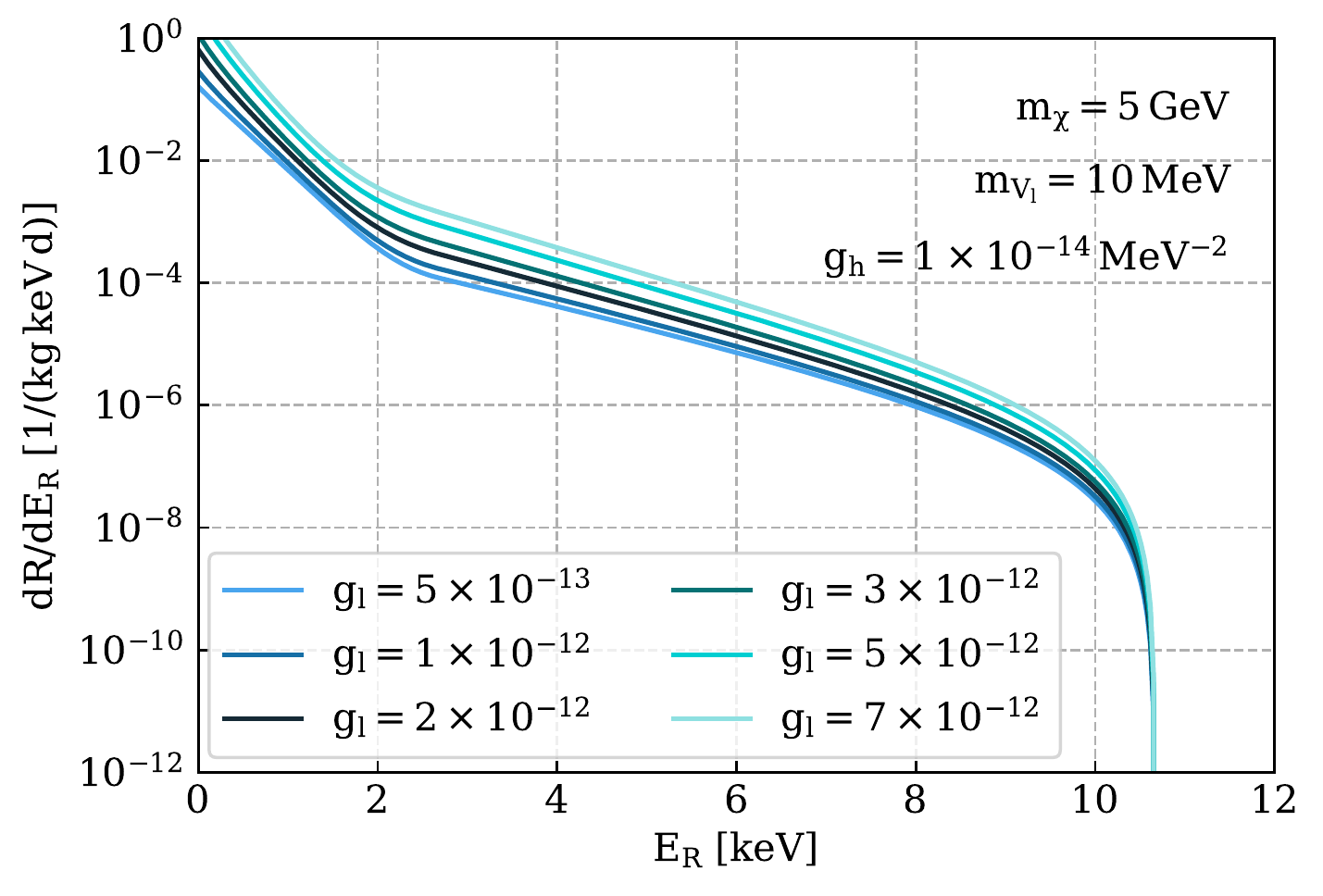}
        \subcaption{}
    \end{subfigure}
    \begin{subfigure}[t]{0.49\textwidth}
        \includegraphics[width=\columnwidth]{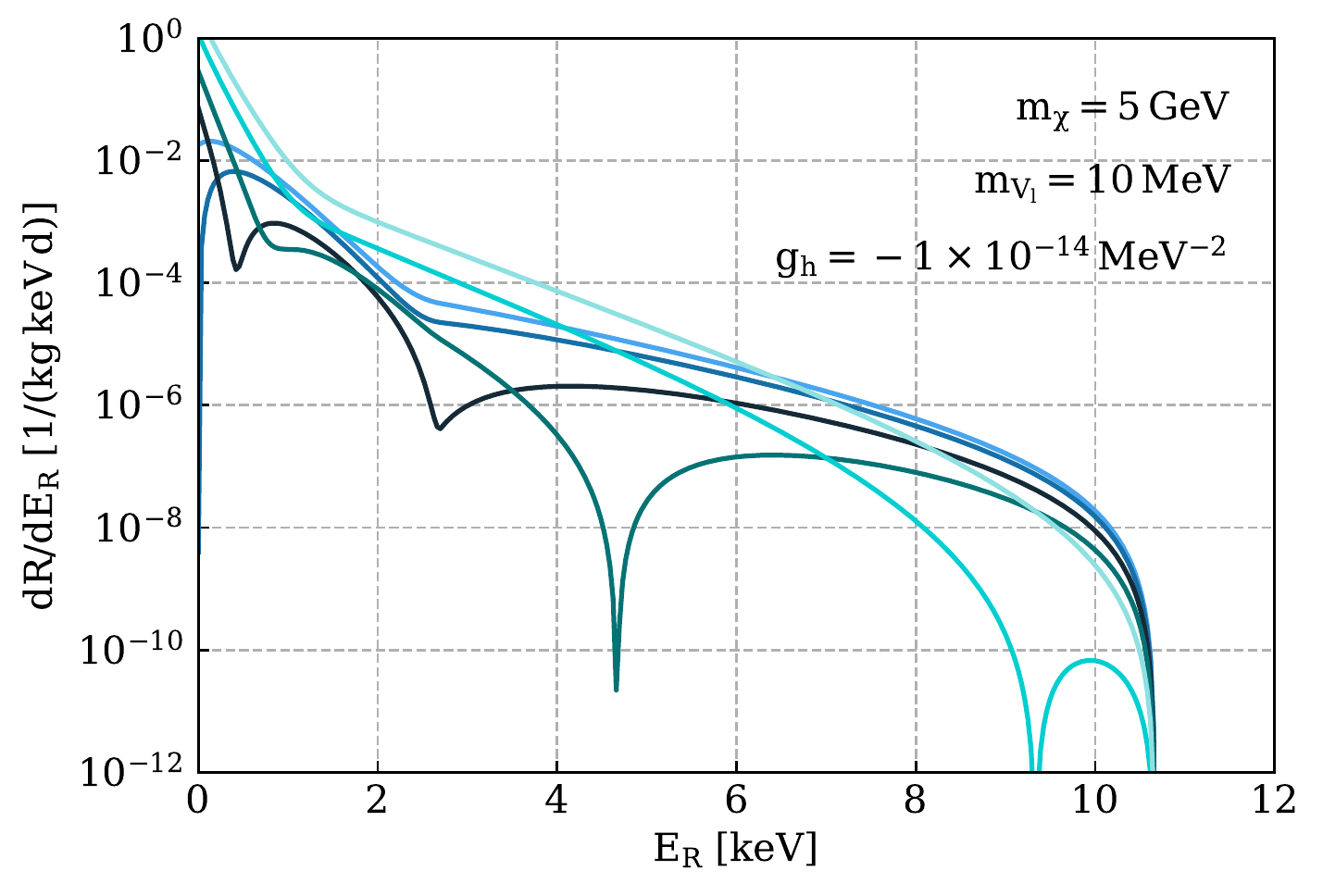}
        \subcaption{}
    \end{subfigure}
    \caption{Differential event rates for the biportal mediator model on a NaI target for the constructive (a) and destructive cases (b). The light mediator mass is fixed at 10 MeV, the DM mass is fixed at 5 GeV, and the effective coupling of the heavy mediator is fixed at $1\times 10 ^{-14}\,\,\rm MeV^{-2}$. Each line shows the spectrum for a different value of the light mediator coupling $g_l$. Neither the detector threshold nor the energy resolution are considered.}
    \label{fig:NaI_gl_dep}
\end{figure}

In the next step, we investigate the impact of the choice of couplings in the biportal model. In order to do this in a systematic way, the effective heavy mediator coupling is fixed while we vary $g_l$. This analysis allows us to study the limiting cases of coupling hierarchies corresponding to either single heavy or single light mediator scenarios and, more importantly, to identify interesting ranges of coupling hierarchies which lead to distinct biportal model phenomenology. In Fig.~\ref{fig:NaI_gl_dep}, the left panel shows the constructive case on the NaI target. For the highest value of the light coupling (light turquoise line), the light mediator dominates the interaction and the spectrum thus looks like that for a single light mediator. On the other hand, for the smallest value of $g_l$ the heavy mediator is the dominant portal and the spectrum thus resembles the single heavy mediator case, including the characteristic kink for NaI at a recoil energy of around 2 keV. In the destructive case shown in the right panel of Fig.~\ref{fig:NaI_gl_dep} one can observe the same limiting cases of the spectra for the highest and lowest absolute values of $g_l$. However, for all other values the interference characteristics are visible. In order for a dip to appear in the biportal spectrum, the individual spectra of the single light and single heavy mediator components need to overlap or lie very close to each other, corresponding to complete or almost complete effacement of the total rate. In the case of smaller contributions to the interaction from the light mediator, such an overlap or approach can only take place at low recoil energies, where the light mediator component shows an enhanced rate.  For lower absolute values of $g_l$, dips thus only appear in the iodine-dominated part of the recoil spectrum close to the threshold. If $g_l$ is increased, dips also start to show in the sodium-dominated area (dark blue line). For higher absolute values of $g_l$, a possible overlap of rates is only achievable at higher recoil energies outside the sodium-dominated regime. This dip gets more pronounced with further increased coupling of the light mediator and moves towards higher energies. At some point, the contribution to the differential recoil rate from the light mediator is much larger than that from the heavy mediator for all recoil energies. 
 
\begin{figure}[tb]
\centering
\includegraphics[width=0.77\columnwidth]{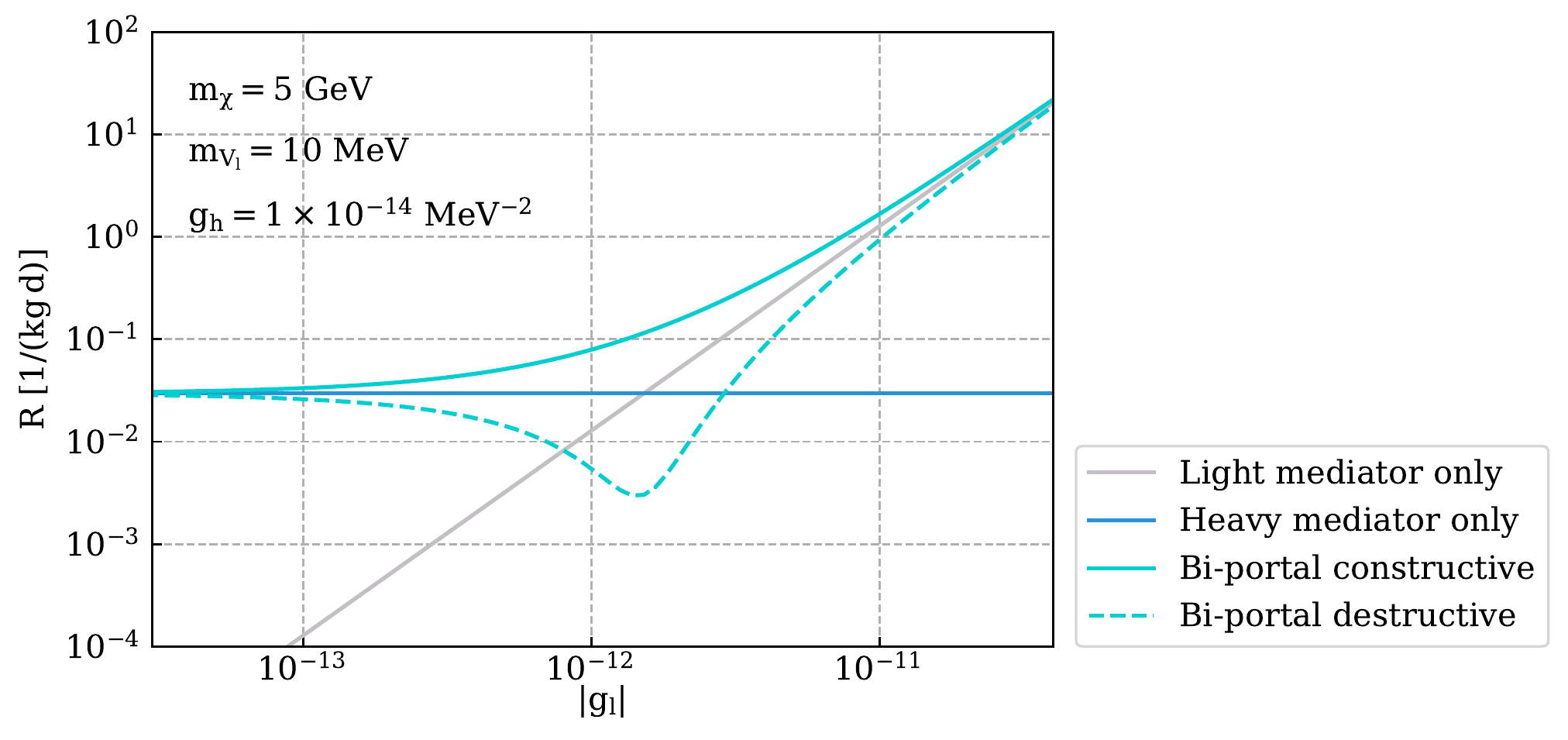}%
\caption{Total event rates on an NaI target as a function of the light mediator coupling. Event rates are shown for single light mediator (grey), single heavy mediator (blue), and biportal models (turquoise) for both constructive (solid lines) and destructive interference (dashed lines). The DM particle mass is fixed at 5 GeV, the light mediator mass is fixed at 10 MeV, and the effective coupling for the heavy mediator is set to $1\times 10^{-14}\,\,\rm MeV^{-2}$. Neither the detector threshold nor the energy resolution effect are considered.}
\label{fig:NaI-totevents}
\end{figure}

In addition to differential recoil spectra, the total number of expected events also offers a tool to compare the phenomenological predictions of various DM-SM interaction models with the experimental data. The total rate per unit exposure is given by
\begin{equation}
    \label{eq:totrate}
    R=\int_{E_{\rm thr}}^{E_{\rm max}} \frac{\d R(E_R^\prime)}{\d E_R}\, \d E_R^\prime,
\end{equation}
where $E_{\rm thr}$ is the energy threshold of the experiment and $E_{\rm max}$ is the maximum energy of the recoil spectrum. In the biportal model, understanding the total event rate as a function of the light mediator coupling for a fixed value of $g_h$ is especially instructive. Setting for the moment $E_{\rm thr}=0$, the predictions for the total rate $R(|g_l|)$ are shown in Fig.~\ref{fig:NaI-totevents} for the biportal model (turquoise; constructive as solid, destructive as dashed lines), as well as for a single light (grey) and a single heavy (blue) mediator model. Showing the total number of events as a function of $g_l$ results in a constant rate for the single heavy mediator case and a rate scaling linearly with the squared coupling for the single light mediator case, respectively. As previously seen for the differential event rate, for low values of $g_l$ the total event rate predicted by the biportal model coincides with the single heavy mediator case. On the other hand, for high values of $g_l$ the light mediator dominates the interaction. While the constructive case is characterized by a smooth transition between the two limiting cases in Fig.~\ref{fig:NaI-totevents}, the destructive case predicts a significantly reduced number of total events for certain values of $g_l$. This suppression takes place for precisely those values of the light mediator coupling for which the destructive interference effects are enhanced for the given choice of $g_h$. Moreover, in the destructive case the total event rate as a function of the light mediator coupling $R(|g_l|)$ is not injective. In other words, a certain number of events can be expected for different combinations of the two couplings. This is an important feature of the destructive biportal model and will play a role later in the discussion of exclusion limits.

\subsection{Impact of detector-specific quantities}
\label{subsec:detectorspefifications}

Detector-specific parameters, such as threshold or energy resolution, were not considered in the previous subsection. However, these are important features of direct-detection experiments and have to be taken into account whenever a statement is made about the sensitivity of DM searches to the model parameter space. In the following we investigate the impact of the detector energy resolution on differential recoil spectra off NaI, as well as on total event rates. We model the finite energy resolution of a detector by a Gaussian distribution of width $\sigma$. The differential event rate is then convolved with this Gaussian distribution. 

\begin{figure}[tb]
\centering
\includegraphics[width=0.7\columnwidth]{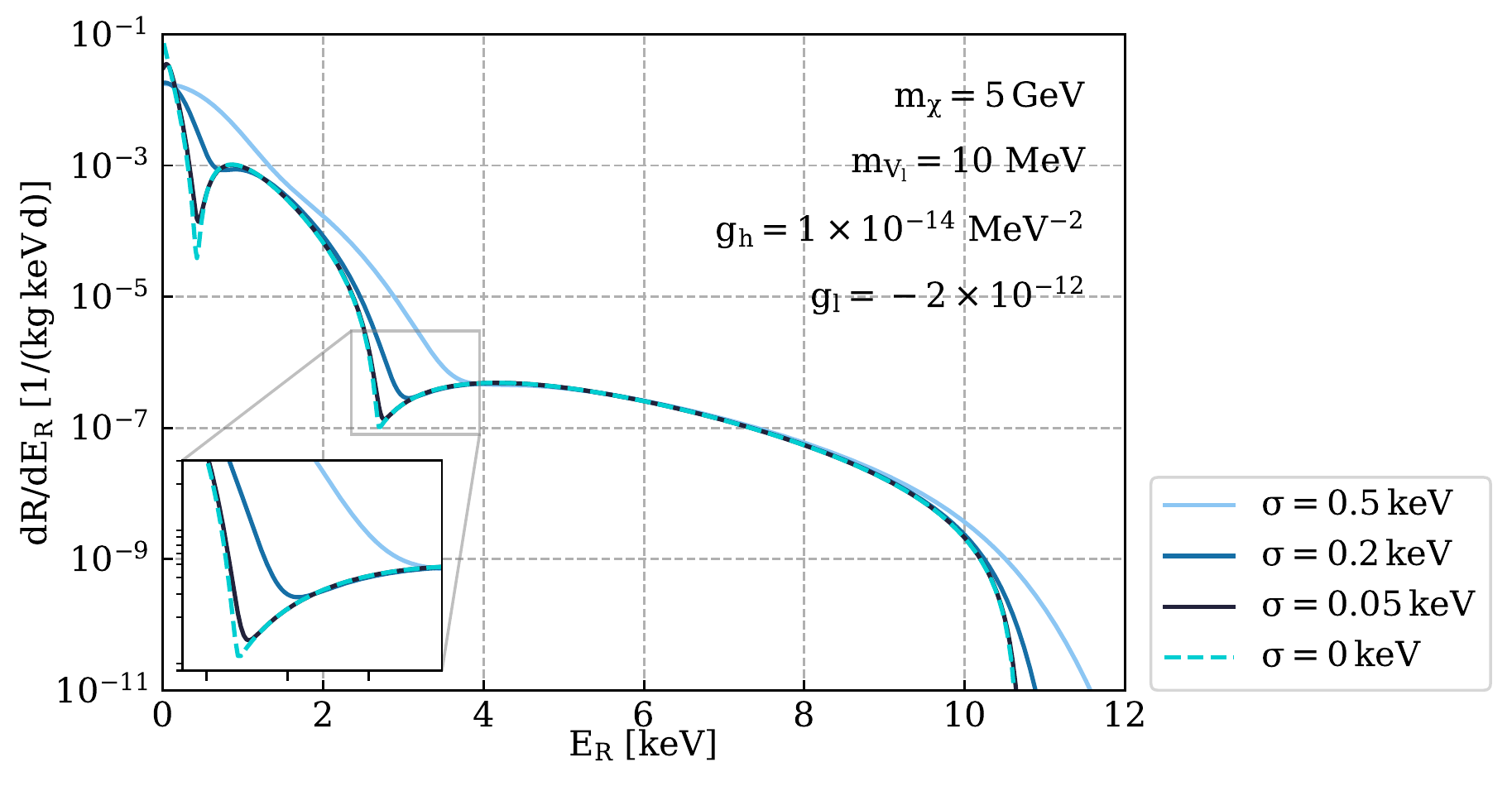}%
\caption{Impact of detector resolution on the destructive interference case of the biportal model on a NaI target. The mass of the light mediator is fixed at 10 MeV, the DM mass is fixed at 5 GeV, the effective coupling of the heavy mediator is fixed at $10^{-14} \,\,\rm MeV ^{-2}$ and the light mediator coupling is fixed at $-4\times 10^{-12}$. Different lines represent different values of resolution, including the reference case of an ideal detector with infinite resolution (turquoise  dashed line).}
\label{fig:NaI-resolution}
\end{figure}

In Fig.~\ref{fig:NaI-resolution} we demonstrate the impact of this convolution with the resolution on the shape of the differential event rate for the destructive interference case in the biportal model on a NaI target. The masses are set equal to their benchmark values, the effective coupling of the heavy mediator to $10^{-14}\,\,\rm {MeV}^{-2}$ and $g_l=-4\times 10^{-12}$. The various lines correspond to different values of the resolution $\sigma$, while the turquoise dashed line shows the spectrum without finite resolution. Poor detector resolutions, i.e.~high values of $\sigma$, can wash out the destructive interference features significantly. However, for high-resolution experiments with values of $\sigma\ll0.05\,\,\rm keV$, such as CRESST-III,\footnote{CRESST-III reaches resolutions as low as 4.6 eV with a CaWO\textsubscript{4} target crystal \cite{CRESST2019_firstresults}.} measurement of the interference dips is in principle possible. Moreover, one has to consider that resolution and threshold are not two independent quantities, but in fact are closely related.  The threshold of an experiment is usually around 4--6 times the value of the resolution $\sigma$ \cite{mancuso_low_2018}. In addition to the flattening of peaks, the corresponding threshold thus might cause an interference feature to lie completely out of the sensitive region of a detector. While this might be the case for the first feature in the spectrum shown in Fig.~\ref{fig:NaI-resolution} for the COSINUS experiment with a  projected resolution of 0.2~keV, the second one is still within reach. A change of couplings might also reposition the first destructive feature in an energy region accessible to COSINUS. 

\begin{figure} [tb]
    \centering
    \includegraphics[width=0.81\columnwidth]{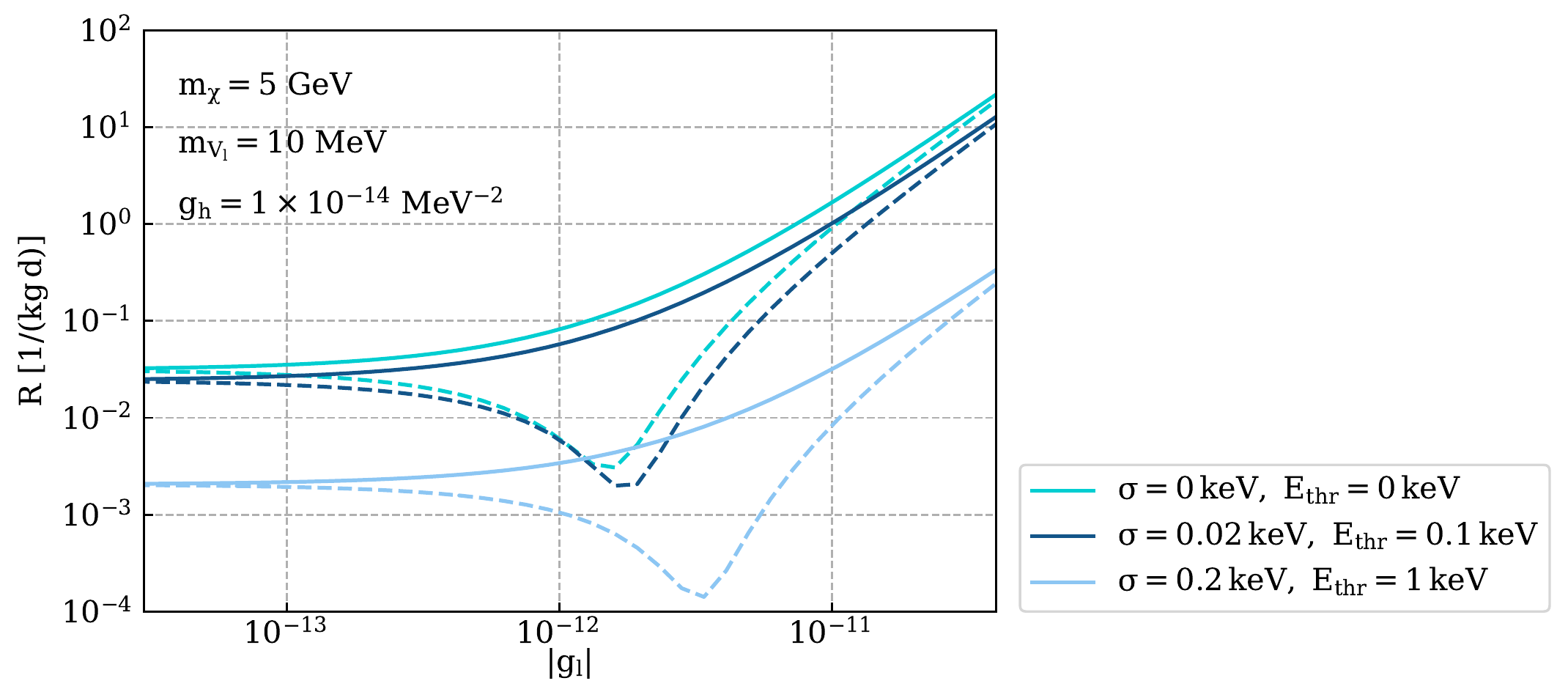}%
    \caption{Total event rates on an NaI target as a function of the light mediator coupling for the biportal model for the constructive (solid lines) and destructive cases (dashed lines). The DM mass is fixed at 5 GeV, the light mediator mass is fixed at 10 MeV, and the effective coupling for the heavy mediator is set to $1\times 10^{-14}\,\,\rm MeV^{-2}$. Different colored lines correspond to different values of the detector resolution. For each resolution an appropriate threshold of $E_{\rm thr}=5\times\sigma$ was set.}
    \label{fig:NaI-totevents2}
\end{figure}

The impact of the detector resolution on the expected total event rate as a function of $g_l$ is shown in Fig.~\ref{fig:NaI-totevents2} for both constructive (solid lines) and destructive (dashed lines) scenarios. Three different cases of detector resolution corresponding to the ideal case (turquoise), 0.02 keV (dark blue), and 0.2 keV (light blue) are considered when computing the total event rate. Our results show that the constructive case is less affected by the detector resolution than the destructive case. For a poor detector resolution, the total event rate for the destructive biportal model can be significantly reduced. As mentioned above, there is a close connection between threshold and resolution. Therefore, as the resolution worsens, the threshold increases, effectively discarding the exponential rise at low recoil energies. Accounting for both of these effects in Fig.~\ref{fig:NaI-totevents2} leads to a further reduction of the overall rate for poor detector resolution. Although the integration is carried out from the threshold over the whole available spectrum, one should note that due to the overall exponentially falling spectrum, the number of total events is dominated by the shape of the spectrum within the first few keV of recoil energy. A higher threshold can thus significantly lower the number of total events. The vertical distance between the various lines in Fig.~\ref{fig:NaI-totevents2} grows slightly with increasing $|g_l|$ as the impact of the threshold is stronger for interactions that are dominated by light mediators. This can be explained by the fact that the differential recoil rates for light mediator interactions are enhanced at low recoil energies, and those low-energy events are then cut off by a high threshold. 

Before proceeding further, a word about the shape of the event rate is in order.
The interference contributions play a significant role only away from the limiting cases where either the light or the heavy mediator are dominant. While constructive interference generally increases the rate, in the case of destructive interference, it is possible to obtain a drastic reduction in the total rate. Such a reduction will play an important role in the upcoming discussion.
\section{Profile likelihood analysis setup}
\label{sec:Analysissetup}
In the next step we do not set the free parameters of the biportal model equal to their benchmark values but instead constrain them using results of current experiments and projections for future ones. As pointed out in the previous section, low-threshold experiments with composite targets are are especially sensitive to the features of the the biportal model with mass hierarchy. We thus perform our analysis using data published by the CRESST Collaboration for their best-performing CaWO\textsubscript{4} detector in the first run of the CRESST-III experiment \cite{CRESST2019_firstresults,cresst2020description}, as well as for the future COSINUS experiment \cite{Angloher:2016ooq}. The latter is a logical choice due to its target material NaI, which has proven especially suitable for our study due to the large difference in mass numbers for sodium and iodine. The experiments we consider and the corresponding parameters are summarized in Table~\ref{tab:expt}.

We use a profile likelihood approach (similarly to e.g., Ref.~\cite{kahlhoefer_exploring_2017}) to calculate the exclusion limits, where the test statistic, as a function of the free parameters of interest $\boldsymbol{\mu}\subset \lbrace m_\chi,m_{Z^\prime},g_l,g_h \rbrace $, is given by 
\begin{equation}
        t_{\boldsymbol{\mu}}=-2\log\left(\displaystyle\frac{\mathcal{L}(\boldsymbol{\mu},\doublehat{l_b})}{\mathcal{L}(\hat{\boldsymbol{\mu}},\hat{l_b})}.\right)
        \label{eq:ratio}
\end{equation}
Here $l_b$ denotes the background level, which is set to the value that maximizes the likelihood function $\mathcal{L}$ with respect to the given value of $\boldsymbol{\mu}$ in the numerator. The expression $\doublehat{l_b}$ is called the conditional maximum likelihood estimator of $l_b$. In the denominator, $\mathcal{L}$ is maximized both in $\boldsymbol{\mu}$ and $l_b$, and so both $\hat{\boldsymbol{\mu}}$ and $\hat{l_b}$ are their unconditional maximum likelihood estimators. The background level is the only nuisance parameter over which we profile in this simplified approach. Hence, astrophysical quantities (like the galactic escape velocity) and detector specific quantities (like the threshold) are assumed to be precisely known. 

A test statistic defined as in Eq.~(\ref{eq:ratio}) follows a $\chi^2$ distribution of degree $n=\dim (\boldsymbol{\mu})$~\cite{Cowan_2011}. One can thus exclude a specific set of parameters for the model at confidence level $1-p$ if
\begin{equation}
    \label{eq:limitcondition}
    t_{\boldsymbol{\mu}} > \rm CDF^{-1}_{\chi^2}(n,1-p),
\end{equation}
where $\rm CDF^{-1}_{\chi^2}(n,x)$ is the inverse of the cumulative distribution function (quantile) for a $\chi^2$ distribution with $n$ degrees of freedom.

\begin{table}[t]
    \small
    \centering
    \renewcommand{\arraystretch}{1.5}
    \begin{tabular}{|>{\centering\arraybackslash}m{4cm}|>{\centering\arraybackslash}m{6cm}|>{\centering\arraybackslash}m{4cm}|} 
     \hline
      & CRESST-III & COSINUS \\
     \hline\hline
        Target Material & CaWO$_4$ & NaI \\
        Net exposure $\epsilon$ & 3.46 kg day & 50 kg day \\
        Threshold $E_{thr}$ & 0.0301 keV & 1 keV \\
        Resolution $\sigma$ & 0.0046 keV & 0.2 keV \\
        Background & 
        $\displaystyle\frac{dR_b}{dE_R} = d + \frac{A}{\tau} \exp{\Big({-\fr{E}{\tau}}\Big)} \frac{1}{\rm{keV\,kg\,d}}$ & $\displaystyle\frac{dR_b}{dE_R} = 1\, \frac{1}{\rm{keV\,kg\,d}}$\\ 
     \hline
    \end{tabular}
    \caption{Summary of experimental parameters used throughout this work. The constants in the CRESST-III background function are derived using data.}
    \label{tab:expt}
\end{table}  

CRESST and COSINUS are being (will be) operated at the Gran Sasso underground laboratory LNGS. In the case of CRESST, these low background levels, together with a gross exposure of $5.594\, \rm kg\,\,days$ (net exposure of $3.46\,\, \rm kg\,\,days$), lead to a lower number of resulting events, allowing us to work with an unbinned likelihood function. This is advantageous, as computational artifacts due to the choice of binning can be avoided. We thus use the following extended likelihood function, to account for the fact that the expected number of events $N_{\rm sample}$ in a measured energy sample $\lbrace E_1, E_2, \dots , E_{N_{\rm sample}} \rbrace$ is itself a random variable:
\begin{equation}
    \label{eq:likelihood}
    \mathcal{L}(\boldsymbol{\mu},l_b) = \displaystyle\frac{\nu(\boldsymbol{\mu},l_b)^{N_{\rm sample}}}{N_{\rm sample}!}e^{-\nu(\boldsymbol{\mu},l_b)}\prod_{i=1}^{N_{\rm sample}} \frac{f(E_i|\boldsymbol{\mu},l_b)}{\nu(\boldsymbol{\mu},l_b)}=\frac{e^{-\nu(\boldsymbol{\mu},l_b)}}{{N_{\rm sample}}!}\prod_{i=1}^{N_{\rm sample}} f(E_i|\boldsymbol{\mu},l_b)
\end{equation}
In the above expression $f(E_i|\boldsymbol{\mu},l_b)$ denotes the non-normalized energy distribution of a certain DM model as measured by a certain detector, and 
\begin{equation}
    \label{eq:Ntot}
    \nu(\boldsymbol{\mu},l_b) = \int_{E_{\rm thr}}^{E_{\rm max}}\d E f(E|\boldsymbol{\mu},l_b) 
\end{equation}
is the total number of expected events that one achieves when integrating the distribution from the energy threshold $E_{\rm thr}$ to the maximum energy $E_{\rm max}(m_\chi)=2\mu_N^2v^2/m_N$. For the projection for COSINUS, the exposure and thus the number of expected events are significantly higher, calling for a binned likelihood approach to reduce the computation time. Thus the data have to be arranged in $N_{\text{bins}}$ bins with $(n_1,\dots,n_{N_{\text{bins}}})$ entries, where the bin  width  $w_\text{bin}$ has to be carefully attuned to the threshold and resolution of the experiment. The likelihood function is then given by
\begin{equation}
    \label{eq:binnedlikelihood2m}
    \mathcal{L}(\boldsymbol{\mu},l_b) = \prod_{i=1}^{N_{\text{bins}}} \frac{\nu_i(\boldsymbol{\mu},l_b)^{n_i}}{n_i!}e^{-\nu_i(\boldsymbol{\mu},l_b)}
\end{equation}
where 
\begin{equation}
    \label{eq:binnedn2m}
    \nu_i(\boldsymbol{\mu},l_b) = \int_{E_\text{\rm thr}+(i-1)w_\text{bin}}^{E_\text{\rm thr}+i w_\text{bin}} f(E|\boldsymbol{\mu},l_b)\d E_R.
\end{equation}

When finding an expression for $f(E_i|\boldsymbol{\mu},l_b)$, the finite energy resolution of the detector is taken into account by convolving the energy distribution as predicted by some DM model, with a Gaussian of width $\sigma$, the energy resolution of the detector. In general the detector energy resolution is an energy-dependent quantity. However, we assume that it is constant, as this is a valid simplification especially at the low recoil energies considered in this analysis. The final expression for the energy distribution thus amounts to
\begin{equation}
    \label{eq:energydistribution}
    f(E|\boldsymbol{\mu},l_b) = \underbrace{\varepsilon \times\int_0^\infty \d E^\prime {\frac{\d R_\chi}{\d E}(\boldsymbol{\mu},E^\prime)}\frac{1}{\sigma\sqrt{2\pi}}e^{-\frac{1}{2}\left(\frac{E-E^\prime}{\sigma}\right)^2}}_{f_\chi(E|\boldsymbol{\mu})} +\underbrace{\varepsilon\times l_b\times\frac{\d R_b}{\d E}(E)}_{f_{b}(E)},
\end{equation}
where $\d R_\chi/\d E$ and $\d R_b/\d E$ denote the differential recoil spectra for the DM model and the background model of the detector, respectively, while $\varepsilon$ gives the exposure. Note here that the background component is not convolved with the resolution. The background for each of the experiments is modeled directly from the data and thus any detector-related properties are already considered. In the case of CRESST-III, the background is fitted to the data under the assumption of zero DM signal. The fit was done using a maximum likelihood approach and it was possible to achieve lowest values of the negative log-likelihood with a background containing a flat plus an exponentially decaying component (see also Table~\ref{tab:expt})\footnote{The fit for the background model is performed only once, under the assumption of no dark matter contribution to the data, rather than refitting for every signal hypothesis. Even though it yields slightly stronger limits, this approach is justified, as the exponential excess present in the CRESST data is of sufficiently different shape from a respective DM contribution for a  large region of the considered parameter space.}. For the COSINUS projection the background distribution is fully determined by the flat model, which is used to generate the mock background sample. 

Within the analysis of real data released by the CRESST Collaboration, two more variables are needed in order to describe the expected energy distribution of the detector. The first variable is the signal survival probability $\xi_\chi$, which accounts for the loss of signal events from applied data selection criteria (cuts). Since the data from the CRESST-III experiment is taken from a so-called acceptance region in the light yield versus energy plane,\footnote{In two-channel experiments, plotting the light yield versus the recoil energy of an observed event offers a way to discriminate nuclear recoils from other particle events, such as electron recoils. In the plane of such a light yield plot, one can thus define a region within which nuclear recoils are expected for each target element. A thorough explanation on how the acceptance region for possible DM events was defined for the CRESST-III experiment can be found in \cite{CRESST2019_firstresults}.} the probability of a signal event lying within the acceptance region, $\xi_{\chi,\rm acc}(E)$, has to be considered as well. Hence, for the CRESST-III data set the energy distribution amounts to
\begin{equation}
    \label{eq:energydistributionCRESST}
    f(E|\boldsymbol{\mu},l_b) = \xi_\chi\times\xi_{\chi,\rm acc}(E)\times f_\chi(E|\boldsymbol{\mu})+f_{b}(E|l_b).
\end{equation}

\begin{figure}
    \centering
    \includegraphics[width=0.75\textwidth]{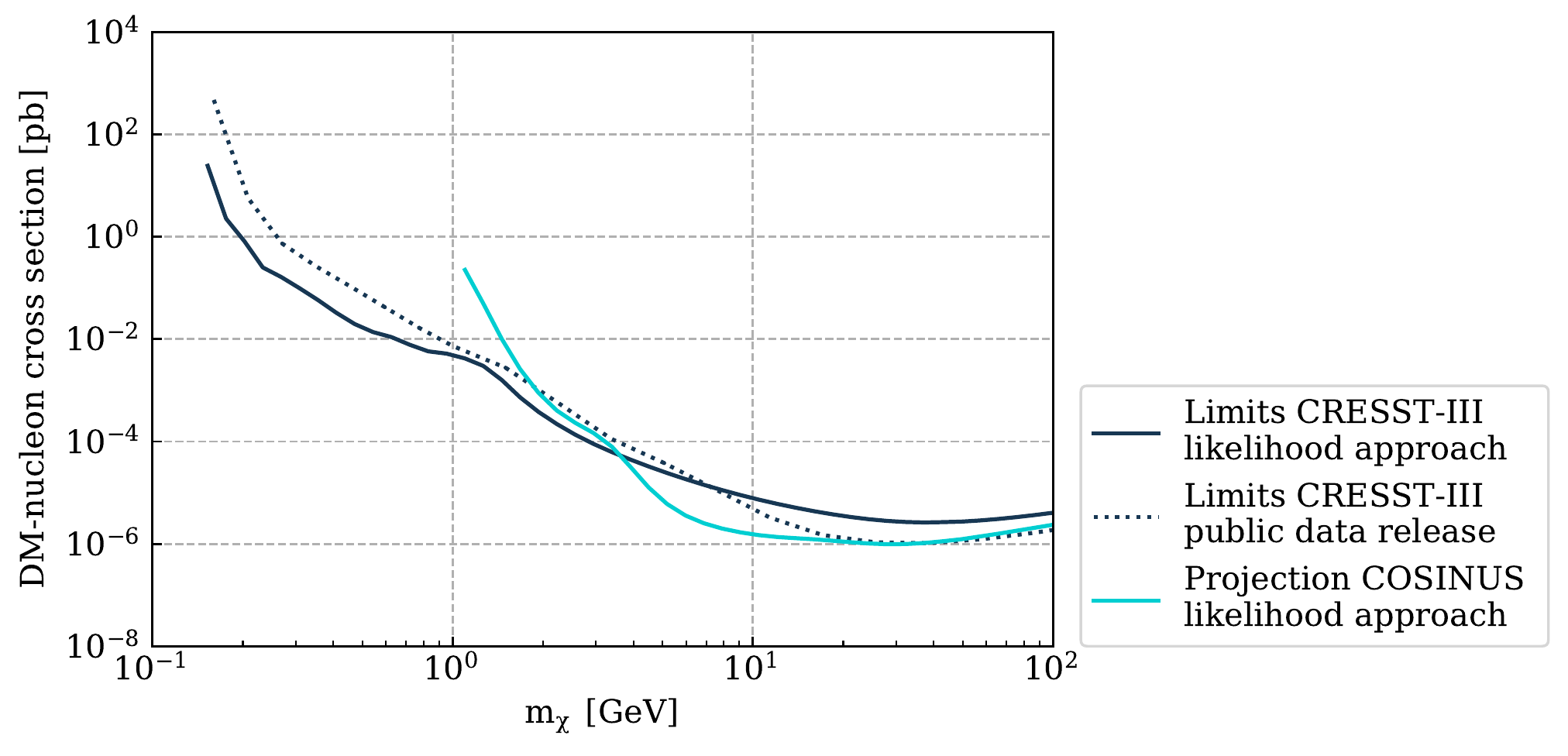}
    \caption{Comparison between 90$\%$ C.L. exclusion limits on the DM-nucleon reference cross section in pb as a function of the DM mass obtained by the CRESST-III experiment (black line) and with our likelihood approach (blue line). We also show the expected sensitivity of COSINUS using our likelihood implementation (cyan line). }
    \label{fig:CRESST_compare}
    \vspace{-0.2cm}
\end{figure}

We validate our approach by rederiving the official CRESST-III exclusion limits. The comparison between our derived limit and the official limit is shown in Fig.~\ref{fig:CRESST_compare}. In general, a good agreement between the two is seen. The small differences in the limit originate from the fact that the limits published by CRESST were calculated with Yellin’s optimum interval method~\cite{Yellin_2002} and we are using a likelihood approach. A main difference between the two is the treatment of the background. The Yellin method makes no assumption on the background and solely exploits differences in the shape of the background and potential signal to set a limit on the cross section. In the likelihood approach we assume a flat background and an exponential for the excess at low energies. Their scaling parameters are the only nuisance parameters of the likelihood. In this way, considering an exponentially decaying background component at low energies leads to stricter limits at low DM masses. Another factor that may lead to some differences between the results is the method with which the expected DM spectrum is determined. In this work this is done by means of the first term in Eq.~(\ref{eq:energydistribution}), while CRESST uses a more complex procedure in their analysis described in Ref.~\cite{CRESST2019_firstresults}. The differences between these two methods were shown in Ref.\cite{cresst2020description}.  Given the differences between the two approaches and a good agreement with the official limits, we consider our approach to be validated and use our implementation to derive limits for the parameters of the biportal model. In addition, we show the expected sensitivity for COSINUS in Fig.~\ref{fig:CRESST_compare}.

\begin{figure}
    \centering
    \includegraphics[width=0.8\textwidth]{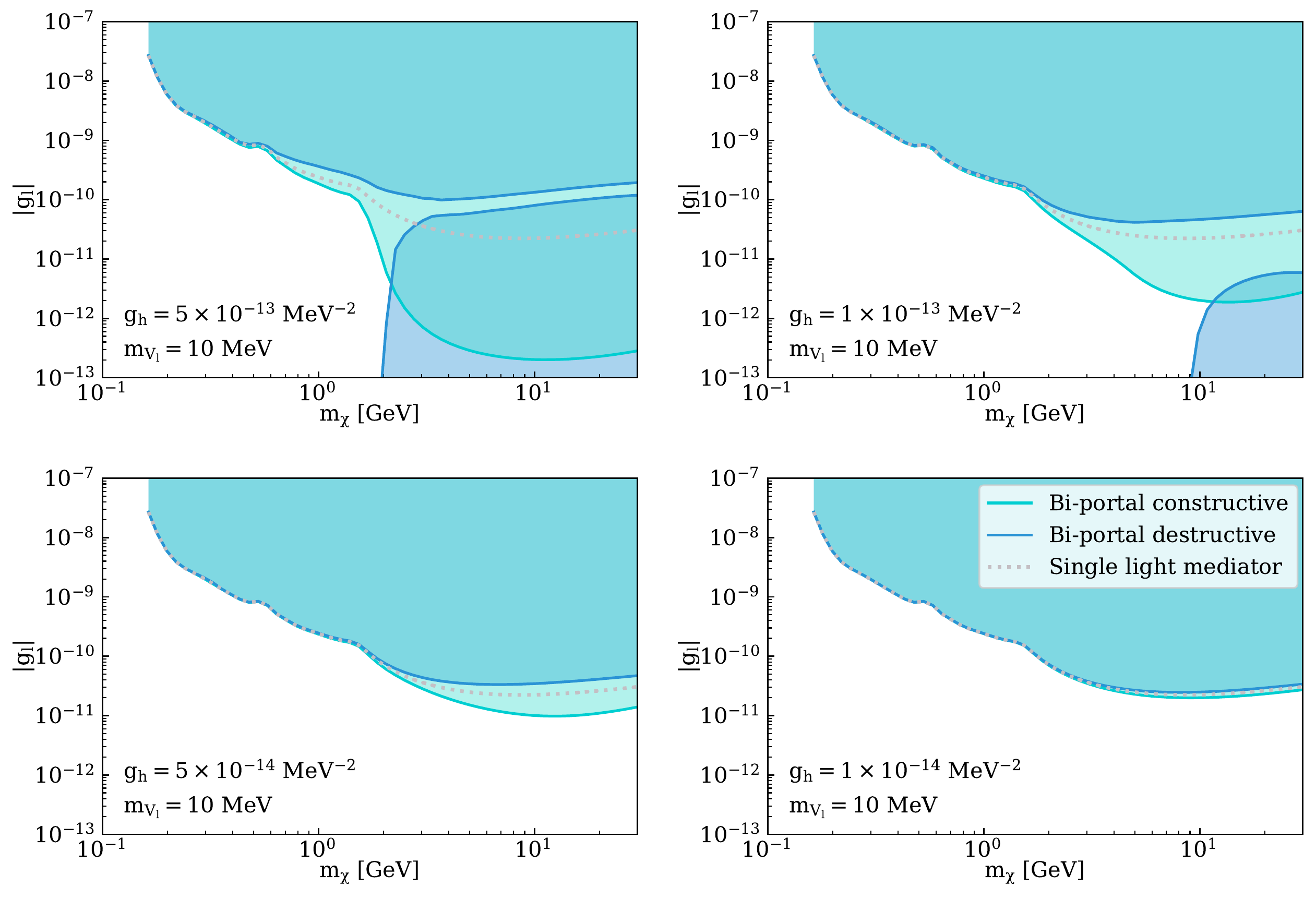}
    \caption{Exclusion limits (at 90$\%$ C.L.) on the light mediator coupling as a function of the DM mass, based on data from the CRESST-III experiment. The light mediator mass is fixed to 10 MeV and each panel shows limits for a different value of the effective heavy mediator coupling. Limits for both the constructive (turquoise lines) and destructive case (blue) are shown. In comparison, we also show limits for a single light mediator case (dashed grey). Shaded regions correspond to excluded parameter space. }
    \label{fig:CaWO4_dmgl_gh}
    \vspace{-0.2cm}
\end{figure}


\section{Results}
\label{sec:results}

We now present our results for the exclusion limits of the four free parameters in the biportal model to which direct-detection event rates are sensitive. In the following, we will assume that the DM relic density constraint is satisfied.\footnote{Unlike the direct-detection experiments, which constrain the product of DM-$V$ and SM-$V$ couplings, the relic density explicitly depends on each of these couplings and the mass of the vector boson $V$.}

There are multiple planes in which the exclusion limits can be drawn, and each of them are instructive. We begin with the analysis in the $m_\chi$-$g_l\,$ plane for the real data from the CRESST-III experiment in Fig.~\ref{fig:CaWO4_dmgl_gh}. While the mass of the light mediator is fixed at 10 MeV in this figure, each panel corresponds to a different value of the heavy mediator coupling. We show the limits for the constructive (turquoise) and destructive (blue) case of the biportal model in comparison with the limit for a single light mediator (i.e. $g_h=0$) (dashed line). The excluded parameter space is shaded in all cases. 
    
Fig.~\ref{fig:CaWO4_dmgl_gh} is best explained by starting with the bottom right panel, where $g_h$ has its smallest value of $1\times 10^{-14}\,\,\rm MeV^{-2}$ and the light mediator thus dominates the interaction process. As expected, the biportal limits in both the constructive and destructive case agree with those for a single light mediator. For a slightly larger heavy-mediator coupling $g_h=1\times 10^{-13}\,\,\rm MeV^{-2}$, displayed in the bottom left panel, the biportal limits start to diverge from the single mediator case. While constructive interference leads to stronger limits -- due to a higher number of predicted events -- the destructive case gives weaker limits. In the upper two panels, the biportal model shows its differences in the allowed parameter space compared to the single-mediator case. For constructive interference the exclusion limits behave as expected: for increasing $g_h$ only lower values of $g_l$ are compatible with the background-only hypothesis and the limits thus get stronger. However, for the destructive case, the nonexcluded region is now not only limited from above, but for certain masses also limited from below.

This behavior can be understood as follows. Concentrating on the top two panels of Fig.~\ref{fig:CaWO4_dmgl_gh}, when $g_l$ is sufficiently large, the total differential cross section is dominated by the light mediator contribution. This together with the threshold of the detector leads to similar limits for the light DM mass around $0.2\,\,\rm{GeV}$ irrespective of the value for $g_h$. As $g_l$ decreases, the total rate decreases, as explained also in Sec.~\ref{subsec:detectorspefifications}. For intermediate values of $g_l$ the destructive interference plays the most important role, thus drastically reducing the total number of events obtained, and hence no limits are obtained. Reducing $g_l$ further renders the contribution of the light mediator negligible; however, sufficient discrimination power is still present in the recoil spectrum due to the heavy mediator contribution $g_h$. This leads to a somewhat artificial upper bound on $g_l$. This is further confirmed by comparing the top two panels of Fig.~\ref{fig:CaWO4_dmgl_gh}, where increasing the heavy mediator coupling $g_h$ leads to a stronger upper bound on $g_l$. 

The parameter space of a single light mediator, biportal constructive and biportal destructive interference models are thus complementary to each other. While the constructive interference biportal model generically leads to stronger limits compared to the single light mediator models, limits from destructive interference biportal scenarios show a different constraining behavior. Generally this leads to two-sided limits and opens up some parameter space which is otherwise disfavored by single light mediator models. We stress that the biportal model studied in this work is capable of testing regions of parameter space that are otherwise not accessible to single-mediator models.

\begin{figure}
    \centering
    \includegraphics[width=0.8\textwidth]{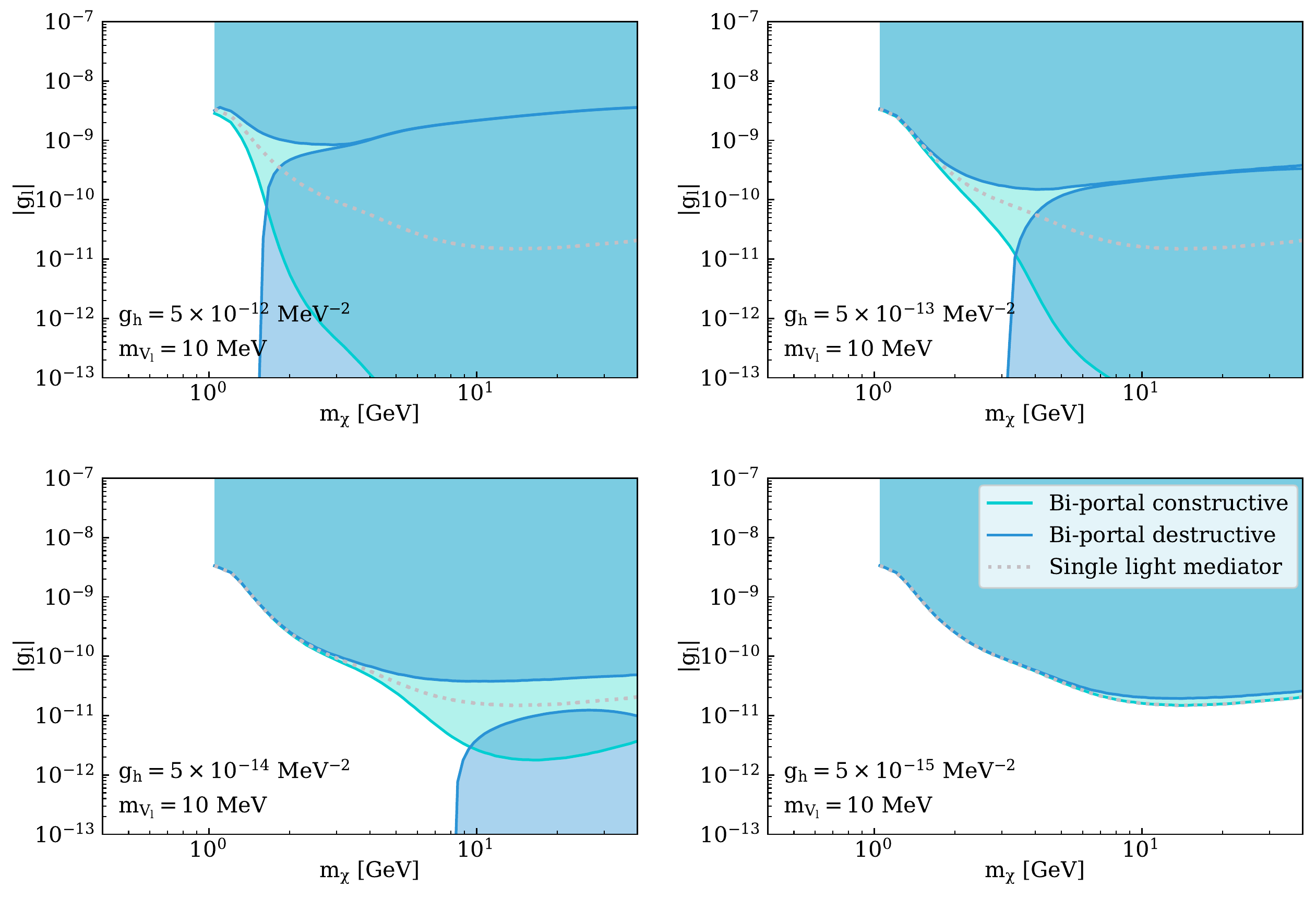}
    \caption{Projected exclusion limits (at 90$\%$ C.L.) on the light mediator coupling as a function of the DM mass, based on mock data for the future COSINUS experiment. The light mediator mass is fixed and each panel shows limits for a different value of the effective heavy mediator coupling. Color codes are the same as those used in Fig.~\ref{fig:CaWO4_dmgl_gh}.}
    \label{fig:NaI_dmgl_gh}
    \vspace{-0.2cm}
\end{figure}

\begin{figure}
    \centering
    \includegraphics[width=0.8\textwidth]{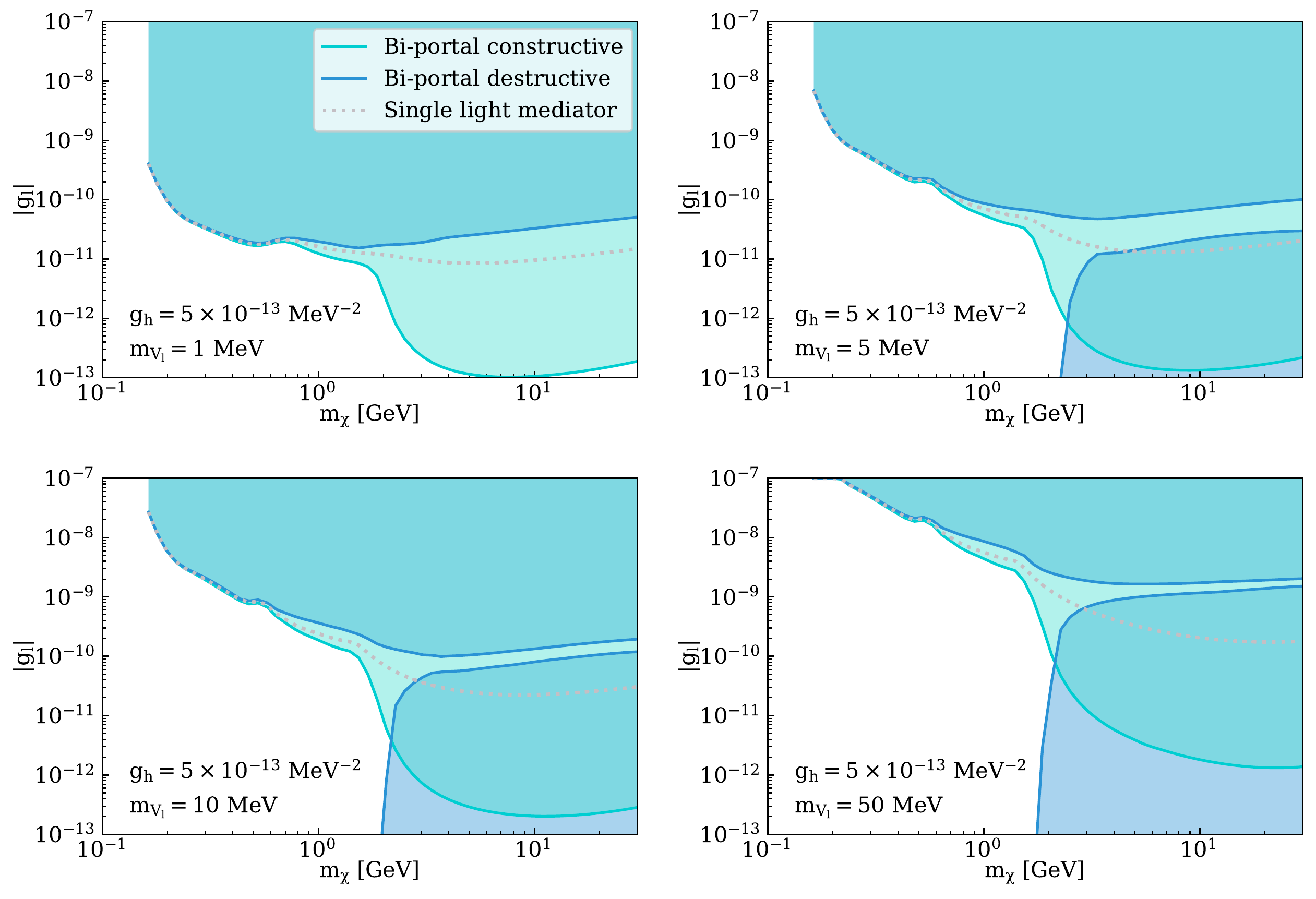}
    \caption{Exclusion limits (at 90$\%$ C.L.) on the light mediator coupling as a function of the DM mass, based on real data from the CRESST-III experiment. The light mediator mass is varied and each panel shows limits for fixed value of the effective heavy mediator coupling of $5 \times 10^{-13}\,\,\rm{MeV}^{-2}$. Color codes are the same as those used in Fig.~\ref{fig:CaWO4_dmgl_gh}.}
    \label{fig:CaWO4_dmgl_mz}
    \vspace{-0.2cm}
\end{figure}

\begin{figure}
    \centering
    \includegraphics[width=0.76\textwidth]{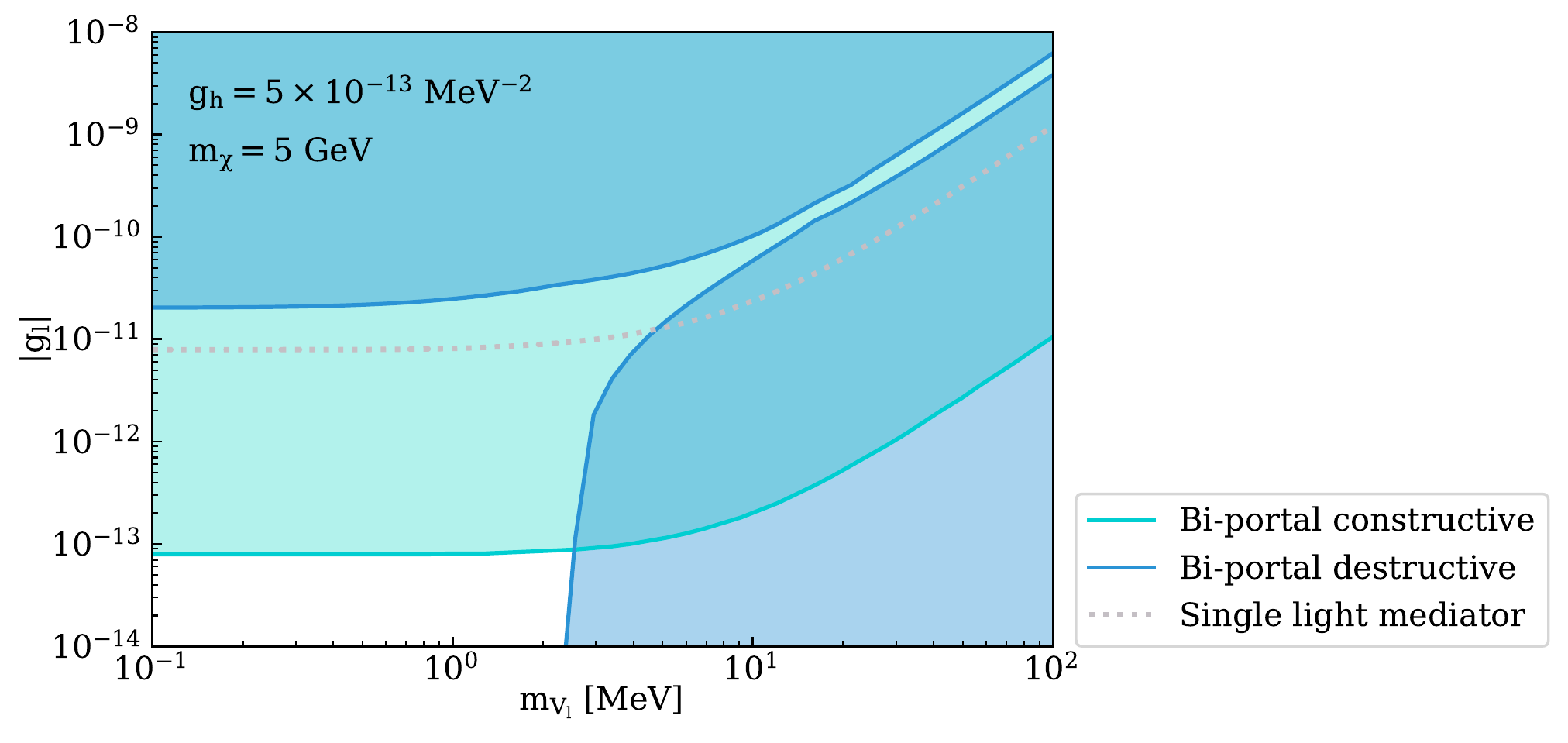}
    \caption{Exclusion limits (at 90$\%$ C.L.) on the light mediator coupling as a function of the light mediator mass, based on real data from the CRESST-III experiment. The DM mass is fixed to 5 GeV and the effective heavy mediator coupling to $5\times 10^{-13}\,\text{MeV}^2$. Color codes are the same as those used in Fig.~\ref{fig:CaWO4_dmgl_gh}.}
    \label{fig:CaWO4_mz_gl}
    \vspace{-0.2cm}
\end{figure}

\begin{figure}
    \centering
    \includegraphics[width=0.8\textwidth]{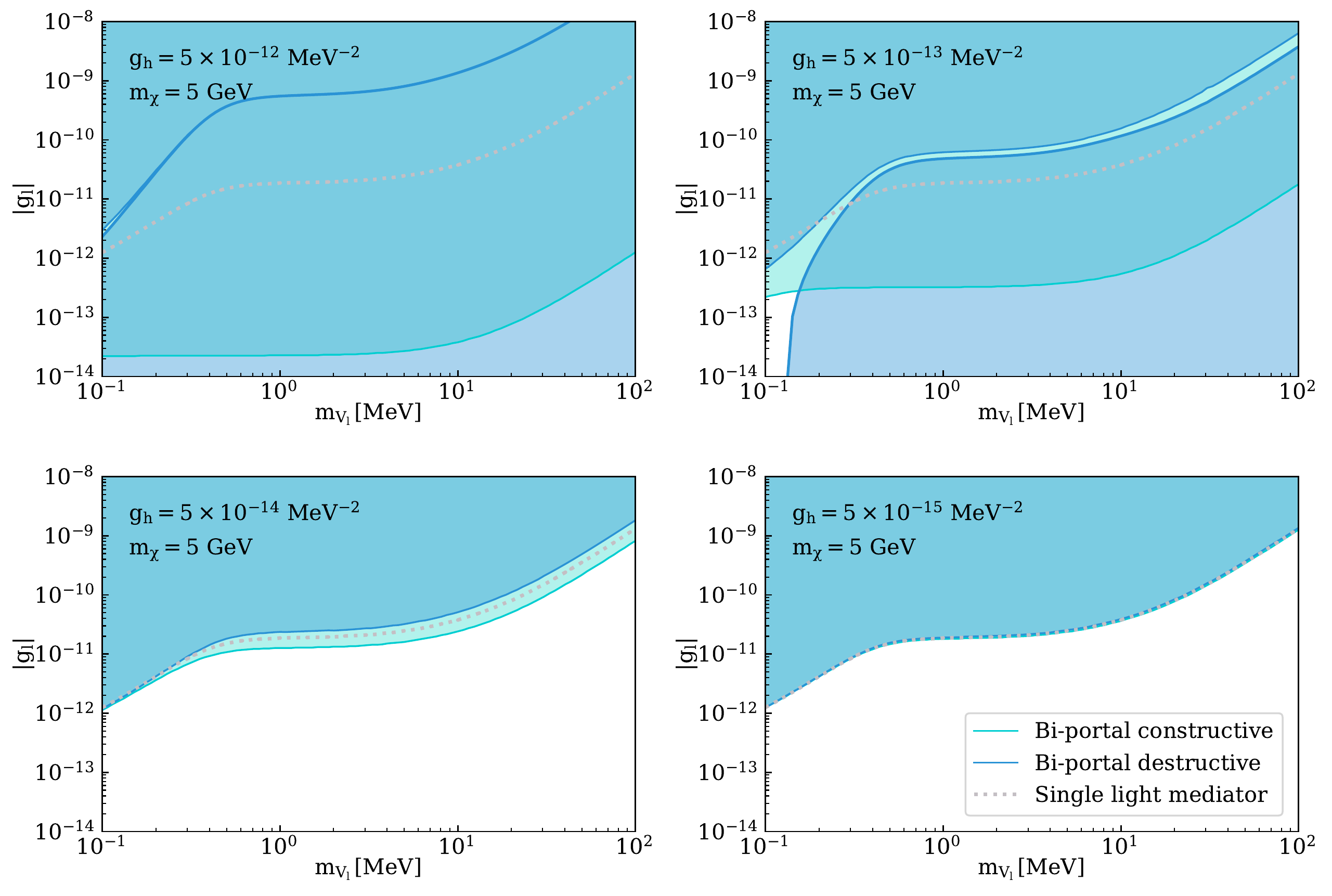}
    \caption{Exclusion limits (at 90$\%$ C.L.) on the light mediator coupling as a function of the DM mass, based on mock data for the future COSINUS experiment. The DM mass is fixed to 5 GeV and each panel shows limits for a different value of the effective heavy mediator coupling. Color codes are the same as those used in Fig.\ref{fig:CaWO4_dmgl_gh}.}
    \label{fig:NaI_mZgl_gh}
    \vspace{-0.2cm}
\end{figure}

Turning our attention to the COSINUS experiment, in Fig.~\ref{fig:NaI_dmgl_gh} we show the projected exclusion limits in the $m_{\chi}-g_l$ plane for fixed light mediator mass and heavy mediator coupling. Comparing this with the limits obtained for $\rm{CaWO}_4$, the qualitative similarity can be immediately seen. There are however crucial differences which arise from the experimental setup. The first and foremost difference is in the reach of the experiments for low DM masses. Given the low threshold, CRESST probes much smaller DM masses compared to COSINUS. Second, the expected COSINUS exposure is about 14 times larger compared to the CRESST data set used in this work,\footnote{The lower exposure of CRESST is not a general limitation of CRESST, but only true for the CRESST-III data set used in this work which is from a single detector only \cite{CRESST2019_firstresults,cresst2020description}. As is shown in Ref.~\cite{CRESST:2015djg}, exposures on the scale of COSINUS or even higher would be within reach also for the small (24 g), low-threshold CRESST-III detectors.} which would make COSINUS probe much smaller values of $g_l$. Finally, an interesting difference between the two can be observed in the limits for the destructive interference case. In the COSINUS experiment, the allowed parameter space corresponding to the strip between the lower and upper bounds on $g_l$ is much narrower compared to CRESST. 
 
In principle, tungsten would be a very good element, leading to a large mass gap in the composition of the CRESST target, but the recoil spectrum from scattering off this element is mostly below threshold and hence tungsten does not contribute for most of the parameter space, except potentially for very light DM masses.

In Fig.~\ref{fig:CaWO4_dmgl_mz} exclusion limits (at 90\% C.L.) using CRESST real data
sample are again shown in the $m_{\chi}-g_l$ plane. Here the effective coupling of the heavy mediator is fixed, while each panel shows limits in the biportal model for  different values of $\mzl$. Excluded regions are again shaded. Overall, larger mediator masses lead to higher allowed values for $g_l$. This can be explained by the inverse proportionality of the differential recoil rate (and also the total event rate) to the mediator mass. For lower $\mzl$ a light-mediator interaction model predicts a higher number of events and the limit on the coupling thus has to be lower. This also has a strong effect on the biportal model. Especially for $\mzl$ = 1 MeV in the upper left panel, the model is dominated by the light mediator and there are no noticeable visible differences between the constructive and destructive cases. Moreover, since the impact of the interference term is very low compared to the contribution from the light mediator, there is no lower limit on $g_l$ in the destructive case. In the next panel with $\mzl$ = 5 MeV, interference effects start to appear again. For higher DM masses the limits are very similar to those in the next panel for $\mzl$ = 10 MeV, which is identical to the already discussed top-right panel of Fig. \ref{fig:CaWO4_dmgl_gh}. For lower DM masses the limits are however flatter, due to the lower mediator mass. 

In Fig.~\ref{fig:CaWO4_mz_gl}, we show the limits derived using CRESST-III real data in the $\gl-\mzl$ plane for a fixed dark matter mass of 5 GeV. The bounds for the constructive (cyan) and destructive (blue) cases of the biportal model are displayed together with the limit resulting from DM-nucleus interaction via a single light mediator (gray dashed). Higher values for the light mediator coupling are allowed for higher mediator masses. This behavior can be explained by the inverse mass proportionality in all light mediator terms contributing to the matrix element, which describes the DM-nucleus interaction. When going from higher to lower mediator masses, the limits get stronger as the predicted number of events increases. For mediator masses under 10 MeV the limits start to flatten. This is the mediator mass regime for which $\mzl < q^2 = 2m_N\,E_R$ and the DM-nucleus scattering can be described by a long-range interaction. The momentum transfer is then the governing quantity in the denominator of the intermediate vector-boson propagator and the mediator mass is expected to only have an impact at very low recoil energies for which $\mzl \geq 2\, m_N\,E_R$. Also, the threshold of the experiment plays an important role here, as contributions from the strong exponential behavior of long-range interactions to the total rate at low recoil energies get cut off. 
As in the $\gl-m_\chi$ plane above, the constructive case of the biportal model gives stronger limits than the single light mediator case. The behavior of the destructive limit is also consistent with Fig.~\ref{fig:CaWO4_dmgl_gh}, showing an additional bound on the mediator coupling from below for higher mediator masses, where the rate is further reduced and the heavy mediator dominates the process. This results again in a band-like structure of the allowed parameter space.

In Fig.~\ref{fig:NaI_mZgl_gh}, we turn again to the COSINUS experiment, and discuss limits in the $\gl-\mzl$ plane for a fixed DM mass of 5 GeV. This figure shows bounds for various values of the heavy mediator coupling to allow for comparison with the panels in Fig.~\ref{fig:NaI_dmgl_gh}. In the lower two panels one can again observe the single light mediator limiting case, while in the upper panels the overall behavior is comparable to the CRESST case.

\section{Relic density}
\label{sec:relic}

Finally, we briefly provide a discussion about relic density generation mechanisms and comment on the possibility of generating relic density within the parameter space we consider. We should remind ourselves that there are in principle seven free parameters in our theory: $\mzh, \mzl, g_{Qh}, g_{Ql}, g_{\chi\,h},g_{\chi\,l},m_{\chi}.$\footnote{In the direct-detection analysis we reduce the parameter space by setting $g_l = g_{Ql}\times g_{\chi\,l}$, $g_h = g_{Qh}\times g_{\chi\,h}/m_{V_h}$} In order to compute the relic density, we need to set all seven parameters. 

Qualitatively speaking, there are three primary annihilation channels that contribute to the relic density, specifically, when $\mzh \gg m_{\chi}$ as we consider here. The first two are the $s$-channel-mediated $\chi\chi\to f\,\bar{f}$ final states. The third is the $t$-channel diagram $\chi\chi\to V_l\,V_l$, which is proportional to $g^2_{\chi\,l}$. For the $s$-channel diagrams there is a contribution from the interference term between the light and heavy mediators. 

Quantitatively, we implement the model in {\tt Micromegas 5.0.2}~\cite{Belanger:2018ccd} to perform the calculations for the relic density. For completeness, we also include vector couplings to leptons, which we set to $g_{Ql}, g_{Qh}$ for the respective interactions.

For concreteness and ease of comparison, we choose the same benchmarks used previously. In particular, we use $\mzl = 20\,\,\rm{MeV}$. We set $\mzh = 500\,\,\rm{GeV}$, $g_{\chi\,h} = 1, g_{Qh} = 0.02$, such that the resulting effective heavy coupling $g_h = 10^{-14}\,\,\rm{MeV}^{-2}$. We then vary $m_{\chi}, g_{Ql}$. 

\begin{figure}[t]
    \centering
    \includegraphics[width=0.45\textwidth]{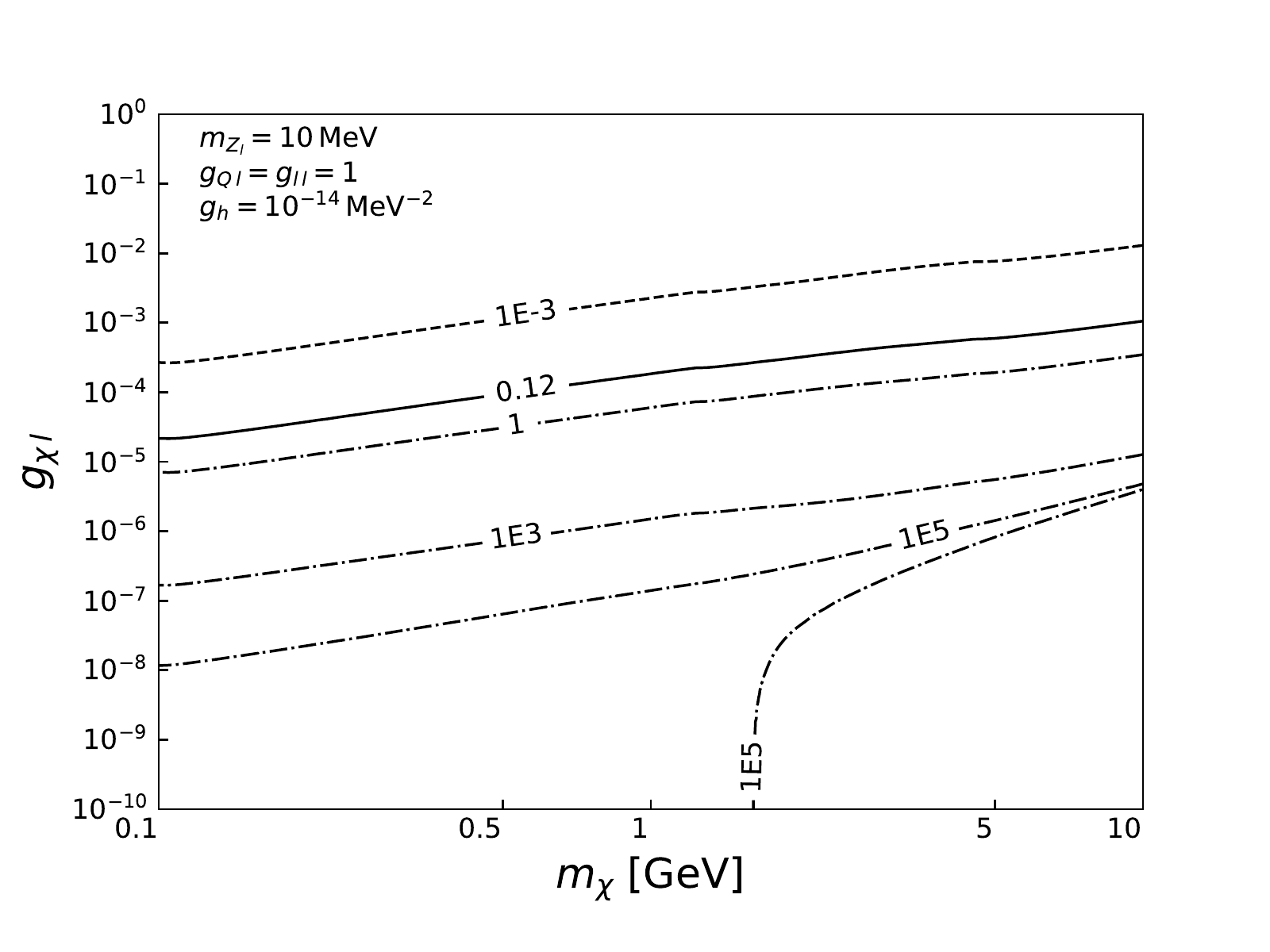}
    \includegraphics[width=0.45\textwidth]{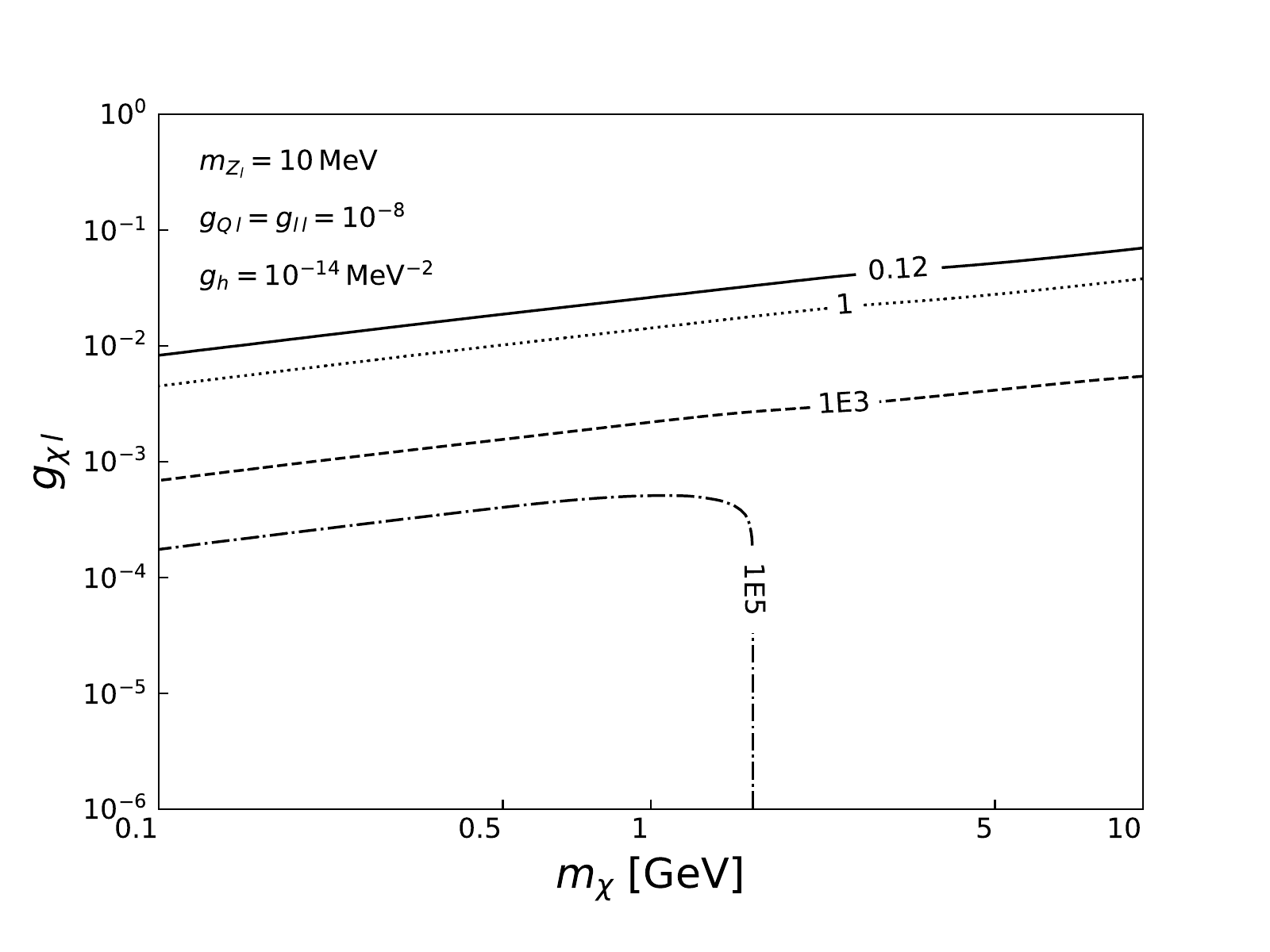}
    \caption{Left: relic density contours in the $m_{\chi}-g_{\chi\,l}$ plane, keeping $m_{V_l} = 10\,\, \rm{MeV}$, $g_h = 10^{-14}\,\,\rm{MeV}^{-2}$, $g_{Ql} = 1$ fixed. Right: relic density contours in the $m_{\chi}-g_{\chi\,l}$ plane, keeping $m_{V_l} = 10\,\,\rm{MeV}$, $g_h = 10^{-14}\,\,\rm{MeV}^{-2}$, $g_{Ql} = 10^{-8}$ fixed. }
    \label{fig:relic}
    \vspace{-0.2cm}
\end{figure}

In the left panel of Fig.~\ref{fig:relic}, we show the resulting relic density contours in the $g_{\chi\,l}$-$m_{\chi}$ plane. When $g_{\chi\,l}$ is large, the $t$-channel diagram contributes and the resulting relic density is driven by the $\chi\chi\to\ V_l\,V_l$ process. Due to the large annihilation cross section in this region, the relic density is very small. As $g_{\chi\,l}$ decreases, the contribution from the $t$-channel process decreases. For smaller values of $g_{\chi\,l}$ the relic density is driven by two $s$-channel diagrams, $\chi \chi \to V_h/V_l\to f\bar{f}$. This cross section is directly proportional to $g^2_{\chi\,l}$, and decreases as $g_{\chi\,l}$ decreases. Correspondingly, the relic density increases.  We obtain the required value of the relic density for $g_{\chi\,l}\sim 10^{-4}$--$10^{-5}$. Decreasing $g_{\chi\,l}$ further leads to overabundant dark matter. 

We also note that there are strong constraints of such light mediators coupling to the SM particles~\cite{Ilten:2018crw}. For 10 MeV dark photons, the strongest constraints limit the dark photon-SM photon mixing parameter $\epsilon \leq 10^{-8}$. Correspondingly, in the right panel of Fig.~\ref{fig:relic} (right panel) we show relic density contours for $g_{Ql} = g_{ll} = 10^{-8}$ with $m_{V_l} = 10\,\,\rm{MeV}$. We observe that for such suppressed couplings to the SM particles, the observed relic density is satisfied for $g_{\chi l} \sim 10^{-2}$. Smaller values of $g_{\chi l}$ typically correspond to overabundant relic density due to suppressed cross sections. 

Finally, we point out that there are several ways in which our results can be made compatible with observations without affecting the conclusions of our previous direct-detection analysis. For example, couplings to leptons could be increased while keeping the ones to quarks unchanged. Furthermore, in a complete theory, there could be additional particles to which dark matter can annihilate, therefore depleting the relic density. For example, introducing axial-vector couplings along with vector couplings would increase the $\chi\chi \to f\bar{f}$ cross section and thus reduce the relic density. Finally, the relic density can be diluted in the early Universe by additional entropy production. 
\section{Conclusions}
\label{sec:conclusions}

Establishing the nature of particle DM and its interactions with SM fields is one of the outstanding open issues in particle physics and cosmology. In the context of direct DM detection, tremendous progress has been recently made on the front of low-threshold, high-resolution experiments, which open up a new testing ground for DM models. Here we presented the case study of a simplified model with one heavy and one light dark vector mediator where interference effects at the level of DM-nucleus scattering amplitudes lead to sharp features in the differential recoil rates. Here, light and heavy are defined with respect to the momentum transfer at direct detection, which depends on the target and the DM mass. Our analysis showed that this kind of model is best explored at experiments with composite targets characterized by a large gap in the atomic mass of their components. Furthermore, we also demonstrated the importance of low-threshold, high-resolution experiments which maximally exploit the information contained in the shape of the recoil spectrum.

We applied a profile likelihood approach to analyze both public real data from the CRESST-III experiment and mock data based on the sensitivity of the future COSINUS experiment and derived limits on the free parameters of our biportal model with vector couplings. We found that CRESST-III already provides strong bounds, which will be further improved by COSINUS projections. To this end, we examined a large region of biportal model parameter space. We showed that CRESST-III public data are in general capable of probing DM masses as low as 0.2 GeV, and constrain light mediator couplings to be lower than $10^{-10}$ for a light mediator mass of 10 MeV. The exact limits on light mediator couplings depend on the strength of the heavy mediator coupling as well. 
With its expected sensitivity the COSINUS experiment can set stronger limits on mediator couplings, albeit looses sensitivity to DM masses less than 1 GeV. This is due to a larger expected exposure, combination of target materials and higher threshold. We note that future CRESST exposures might also be larger, which we have not considered here. Once again, it is important to remember that the exact values of the allowed light mediator couplings depend on the heavy mediator couplings as well. We verified that
constraints on the (overall) coupling weaken for light mediators with masses > 50 MeV.    
In the future, it will be interesting to study further biportal simplified models and UV-complete theories characterized by sizable interference effects e.g., involving axial-vector couplings, and investigate the power of direct-detection experiments in determining viable regions in parameter space. Further complementary studies can involve constraints provided by other DM searches and DM relic density.

Our study provides a specific example of the importance that low-threshold, high-resolution direct-detection experiments can have in guiding model building in particle DM by opening up the possibility to test novel features of recoil spectra and unexplored signatures. 



\paragraph*{Acknowledgments}
S.K. is supported by Austrian Science Fund Elise-Richter grant project number V592-N27. S.K. thanks Pavel Fileviez P\'erez for useful comments. We thank Felix Kahlhoefer for carefully reading the manuscript and useful comments. We thank Federica Petricca for valuable comments on the manuscript.
\bibliography{Refs}
\end{document}